\documentstyle[epsfig]{mn}
\newif\ifAMStwofonts

\ifoldfss
  \ifCUPmtlplainloaded \else
    \NewTextAlphabet{textbfit} {cmbxti10} {}
    \NewTextAlphabet{textbfss} {cmssbx10} {}
    \NewMathAlphabet{mathbfit} {cmbxti10} {} 
    \NewMathAlphabet{mathbfss} {cmssbx10} {} 
  \fi
  \ifAMStwofonts
    \ifCUPmtlplainloaded \else
      \NewSymbolFont{upmath} {eurm10}
      \NewSymbolFont{AMSa} {msam10}
      \NewMathSymbol{\upi}     {0}{upmath}{19}
      \NewMathSymbol{\umu}     {0}{upmath}{16}
      \NewMathSymbol{\upartial}{0}{upmath}{40}
      \NewMathSymbol{\leqslant}{3}{AMSa}{36}
      \NewMathSymbol{\geqslant}{3}{AMSa}{3E}

       \let\le=\leqslant
       
    \fi
  \fi
\fi 

\ifnfssone
  \newmathalphabet{\mathit}
  \addtoversion{normal}{\mathit}{cmr}{m}{it}
  \addtoversion{bold}{\mathit}{cmr}{bx}{it}
  \newmathalphabet{\mathbfit} 
  \addtoversion{normal}{\mathbfit}{cmr}{bx}{it}
  \addtoversion{bold}{\mathbfit}{cmr}{bx}{it}
  \newmathalphabet{\mathbfss} 
  \addtoversion{normal}{\mathbfss}{cmss}{bx}{n}
  \addtoversion{bold}{\mathbfss}{cmss}{bx}{n}
  \ifAMStwofonts
    \ifCUPmtlplainloaded \else
      \UseAMStwoboldmath
      \makeatletter
      \new@mathgroup\upmath@group
      \define@mathgroup\mv@normal\upmath@group{eur}{m}{n}
      \define@mathgroup\mv@bold\upmath@group{eur}{b}{n}
      \edef\UPM{\hexnumber\upmath@group}
      \new@mathgroup\amsa@group
      \define@mathgroup\mv@normal\amsa@group{msa}{m}{n}
      \define@mathgroup\mv@bold\amsa@group{msa}{m}{n}
      \edef\AMSa{\hexnumber\amsa@group}
      \makeatother
      \mathchardef\upi="0\UPM19
      \mathchardef\umu="0\UPM16
      \mathchardef\upartial="0\UPM40
      \mathchardef\leqslant="3\AMSa36
      \mathchardef\geqslant="3\AMSa3E

       \let\le=\leqslant

    \fi
  \fi
\fi 

\ifnfsstwo
  \DeclareMathAlphabet{\mathbfit}{OT1}{cmr}{bx}{it}
  \SetMathAlphabet\mathbfit{bold}{OT1}{cmr}{bx}{it}
  \DeclareMathAlphabet{\mathbfss}{OT1}{cmss}{bx}{n}
  \SetMathAlphabet\mathbfss{bold}{OT1}{cmss}{bx}{n}
  \ifAMStwofonts
    \ifCUPmtlplainloaded \else
      \DeclareSymbolFont{UPM}{U}{eur}{m}{n}
      \SetSymbolFont{UPM}{bold}{U}{eur}{b}{n}
      \DeclareSymbolFont{AMSa}{U}{msa}{m}{n}
      \DeclareMathSymbol{\upi}{0}{UPM}{"19}
      \DeclareMathSymbol{\umu}{0}{UPM}{"16}
      \DeclareMathSymbol{\upartial}{0}{UPM}{"40}
      \DeclareMathSymbol{\leqslant}{3}{AMSa}{"36}
      \DeclareMathSymbol{\geqslant}{3}{AMSa}{"3E}

       \let\le=\leqslant

    \fi
  \fi
\fi 

\ifCUPmtlplainloaded \else
  \ifAMStwofonts \else 
    \def\upi{\pi}
    \def\umu{\mu}
    \def\upartial{\partial}
  \fi
\fi

\def\pd#1#2{{\upartial #1 \over \upartial #2}}
\def\spd#1#2#3{{\upartial ^2 #1 \over \upartial #2 \upartial #3}}

\def\tfrac#1#2{{\textstyle\frac{#1}{#2}}}
\title
[CMB power spectrum estimation]
{Maximum-likelihood estimation of the CMB power spectrum from
interferometer observations}
\author[M.P.~Hobson and Klaus Maisinger]
{M.P.~Hobson and Klaus Maisinger\\ 
Astrophysics Group,
Cavendish Laboratory, Madingley Road, Cambridge, CB3 0HE, UK}
\date{Accepted ---. Received ---; in original form \today}
\pagerange{\pageref{firstpage}--\pageref{lastpage}}
\pubyear{2002}

\begin{document}
\maketitle
\label{firstpage}

\begin{abstract}
  A maximum-likelihood method is presented for estimating the power
  spectrum of anisotropies in the cosmic microwave background (CMB)
  from interferometer observations.  The method
  calculates flat band-power estimates in separate bins in
  $\ell$-space, together with confidence intervals on the power in
  each bin. For multifrequency data, the power spectrum of the other
  foreground components may also be recovered.  Advantage is taken of
  several characteristic properties of interferometer data, which
  together allow the fast calculation of the likelihood function. The
  method may be applied to single-field or mosaiced observations, and
  proper account can be taken of non-coplanar baselines. The method is
  illustrated by application to simulated data from the Very Small
  Array.
\end{abstract}


\section{Introduction}
\label{intro}

Interferometers have proven themselves to be valuable tools in
observing anisotropies in the cosmic microwave background (CMB). They
are inherently insensitive to anisotropies in atmospheric emission on
scales larger than the beam and to ground spillover, and the fact that
they sample Fourier space directly make interferometers ideal
instruments for measuring the CMB power spectrum.

A method for calculating the maximum-likelihood CMB power spectrum
directly from the complex visibility data produced by an
interferometer was first discussed by Hobson, Lasenby \& Jones (1995)
(hereafter HLJ95). The method was illustrated by assuming the CMB
anisotropies to be described by a Gaussian autocorrelation function,
but the technique is easily modified so that the CMB power spectrum is
parameterised in terms of flat band-powers in separate bins in
$\ell$-space.  Indeed, the modified HLJ95 algorithm was used to
calculate flat band-power estimates of the CMB power spectrum in two
spectral bins at $\ell \approx 410$ and $\ell \approx 590$ (with
bin-width $\Delta\ell \approx 90$) for the Cosmic Anisotropy Telescope
(CAT), which is a 3-element interferometer (Scott et al. 1996; Baker
et al. 1999)

Following the early success of the CAT, a new generation of CMB
interferometers have been built, and have recently made
high-sensitivity observations of the CMB.  These experiments include
the Very Small Array (VSA) (Jones 1997; Jones \& Scott 1998), the
Degree Angular Scale Interferometer (DASI) (Leitch et al. 2001;
Halverson et al. 2001) and the Cosmic Background Imager (CBI) (Pearson
et al. 2000; Padin et al. 2001). Although the detailed design of these
experiments is different in each case, the basic principles underlying
their operation are the same.  In particular, these instruments have a
larger number of antennas than the CAT (for example, the VSA has 14
horns) and more sensitive detectors. These specifications enable the
accurate measurement of the CMB power spectrum over a wide range of
angular scales.  For example, the VSA in its `compact' configuration
(see section~\ref{application}) 
measures the CMB power spectrum in 10 independent bins of
width $\Delta\ell \approx 90$ from $\ell \approx 80 - 950$.
 
Some discussion of how to obtain flat band-power estimates of the CMB
power spectrum from this new generation of interferometer experiments
has been presented by White et al. (1999a) and White et al. (1999b),
mainly in connection with the analysis of DASI data. Indeed, the
techniques outlined by these authors have been applied to the analysis
of DASI and CBI observations.  In particular, the DASI experiment has
recently produced an accurate determination of the CMB power spectrum
in 9 spectral bins of width $\Delta\ell \approx 80$ in the range $\ell
\approx 100 - 900$ (Halverson et al. 2001), whereas the CBI has
measured flat band-power estimates of the CMB power spectrum in two
spectral bins at $\ell \approx 600$ and $\ell \approx 1200$, with a
spectral resolution of $\Delta\ell \approx 400$ (Padin et al. 2001).

The increase in both the amount of visibility data and the number of
independent spectral bins means that the computational burden of
performing a likelihood analysis of the new generation of
interferometer experiments is considerable. It is therefore of
interest to investigate fast methods of calculating the likelihood
function for interferometer observations of the CMB.  In this paper,
we extend the discussion given by HJL95 and White et al. (1999a,
1999b) and present a complete description of a maximum-likelihood
technique for calculating flat band-power estimates of the CMB power
spectrum from interferometer data.  Following HJL95, the technique
allows for contributions to the visibilities from Galactic foreground
emission, which can be separated out if multifrequency data are
available. In terms of the computational algorithm, we also present
some straightforward devices for speeding up the calculation of the
likelihood function, which take advantage of certain useful properties
of interferometer data.  Given the facility of fast evaluation of the
likelihood function, we then investigate the relative merits of
obtaining the maximum-likelihood flat band-power estimates, and their
corresponding confidence intervals, either by direct evaluation the
likelihood distribution, or by using
traditional numerical maximisation techniques.

Our formalism makes no special assumptions specific to a particular
experiment, and so can be applied to data from any CMB interferometer,
including observations of mosaiced fields, and proper account is taken
of non-coplanar baselines. The method is illustrated by applying it to
simulated data from the Very Small Array (VSA). The application of the
technique to real VSA observations will be presented in a forthcoming
paper. We note, in passing, that a maximum-entropy map-making method for
interferometer observations, which can
simultaneously deconvolve the interferometer beam and separate
CMB and Galactic foreground emission, is discussed by
Maisinger, Hobson \& Lasenby (1997).

\section{Interferometer observations}
\label{interobs}

Let us consider how an interferometer performs an observation of the
CMB. We begin by discussing the likely contribution to the visibility
data due to emission from different physical components at microwave
wavelengths.

\subsection{Model of the microwave sky}
\label{skymodel}

The total sky intensity $I(\hat{\bmath{x}},\nu)$ at a frequency $\nu$
in a direction $\hat{\bmath{x}}$ will, in general, contain
contributions from the CMB and several foreground components, such as
Galactic free-free and synchrotron emission, as well as emission from
extragalactic radio point sources. As we discuss in
section~\ref{sec:baddatapointsources}, contamination by point source
emission is usually addressed by making simultaneous high-resolution
observations of the sources, which may then be used to subtract the
point source contribution from the data.  We will therefore assume
that point emission has been subtracted in this way, so that the
remaining contamination of the CMB signal is due to diffuse Galactic
emission.

We may express the fluctuations in the sky intensity as a sum over
those due to each of these diffuse components
\begin{equation}
\Delta I(\hat{\bmath{x}},\nu) = \sum_{p=1}^{N_p} \Delta I_{\rm
p}(\hat{\bmath{x}},\nu), 
\label{eqn1}
\end{equation}
where $N_p$ is the number of distinct physical components contributing
to the total sky emission.  For the CMB component, it is more usual to
work in terms of equivalent thermodynamic temperature fluctuations
$\Delta T_{\rm cmb}(\hat{\bmath{x}})$, which is related to the
intensity fluctuations by
\[
\Delta I_{\rm cmb}(\hat{\bmath{x}},\nu) 
\approx \left.\pd{B(\nu,T)}{T}\right|_{T=T_0} 
\Delta T_{\rm cmb}(\hat{\bmath{x}}).
\]
where $B(\nu,T)$ is the Planck function and $T_0=2.726$ K is the mean
temperature of the CMB (Mather et al. 1994). The conversion factor can
be approximated by
\[
\left.\pd{B(\nu,T)}{T}\right|_{T=T_0}
\approx 24.8~\left(\frac{x^2}{\sinh x/2}\right)^2
~\mbox{Jy sr$^{-1}$ ($\mu$K)$^{-1}$},
\]
where $x = h\nu/kT_0 \approx (\nu/56.8~\mbox{GHz})$.  For simplicity
we define the brightness temperature fluctuations for the foreground
components in a similar way. Thus, for the $p$th physical component of
emission, we define
\[
\Delta T_p(\hat{\bmath{x}},\nu) 
= \frac{\Delta I_p(\hat{\bmath{x}},\nu)}
{\partial B(\nu,T_0)/\partial T}.
\]
We note that, in general, the `temperature' fluctuations of the
foreground components will be frequency dependent, unlike those of the
CMB.

Temperature fluctuations on the sky are usually described by an
expansion in spherical harmonics, so for each component we have
\[
\frac{\Delta T_p(\hat{\bmath{x}},\nu)}{T}   
= \sum_{\ell=0}^{\infty} \sum_{m=-\ell}^{\ell}
a_{\ell m}^{(p)}(\nu) Y_{\ell m} (\hat{\bmath{x}}),
\]
from which we define the ensemble-average power spectrum of the $p$th
component at some reference frequency $\nu_0$ by
\[
c^{(p)}_{\ell}(\nu_0) = \langle |a^{(p)}_{\ell m}(\nu_0)|^2\rangle.
\]
Since the power spectrum of the CMB is usually quoted in terms of the
power per unit logarithmic interval in $\ell$, it is also useful to
define the quantities
\[
d^{(p)}_{\ell}(\nu_0) = \ell(\ell + 1)c^{(p)}_{\ell}(\nu_0).
\]

For observations of small patches of sky, however, the spherical
harmonic expansion is awkward to apply, and it is more convenient to
use Fourier analysis, so that
\[
\frac{\Delta T_p(\hat{\bmath{x}},\nu)}{T}
=\int a^{(p)}(\bmath{u},\nu)
\exp(2\upi i\bmath{u}\cdot\hat{\bmath{x}})~{\rm d}^2\bmath{u},
\]
where $a^{(p)}(\bmath{u},\nu)$ is the Fourier transform of the
temperature fluctuations in the $p$th physical component at an
observing frequency $\nu$. The reason for including the factor of
$2\pi$ in the exponent will become clear shortly, when we consider how
an interferometer observes the sky.  The two-dimensional transform
variable $\bmath{u}$ is measured in wavelengths, and for later
convenience we use $\rho=|\bmath{u}|$ to denote radial distances from
the origin in the Fourier domain.  For small fields, $\rho$ is related
to the multipole $\ell$ in the spherical harmonic expansion by $\rho
\approx \ell/2\pi$; the exact relationship is
$\rho=\frac{1}{2}\cot(\pi/\ell)$ (see, for example, Hobson \& Magueijo
1996).  Since Fourier transformation is a linear operation, we may
simply sum over the Fourier modes for each component to obtain an
expression analogous to (\ref{eqn1}) for the Fourier transform of the
total sky fluctuations
\[
a(\bmath{u},\nu) = \sum_{p=1}^{N_p} a^{(p)}(\bmath{u},\nu).
\]

It is convenient to separate the dependence of the Fourier modes on
frequency $\nu$ and on position in the Fourier domain $\bmath{u}$.  We
are thus assuming that the spectral index of any Galactic emission
does is not spatially varying over the observed field. For the compact
VSA, for example, the extent of a single field is $\sim 4.6$ degrees, and so
this assumption is not too severe.  For mosaiced observations 
(see section~\ref{sec:mosaicing}) of
larger areas of sky, however, this assumption may be questionable. We
describe the frequency variation of each component through the
functions $f_p(\nu)$ $(p=1,2,\ldots,N_p)$. In this paper, we take the
reference frequency $\nu_0=30$ GHz and normalise the frequency
dependencies so that $f_p(\nu_0)=1$ for all physical components.

If the emission in each component is statistically isotropic over the
observed field and no correlations exist between the components, then
the Fourier modes have mean zero and their covariance properties are
given by
\[
\langle a^{(p)}(\bmath{u},\nu) [a^{(p')}(\bmath{u}',\nu')]^* \rangle  =  
\]
\begin{equation}
\qquad\qquad\qquad C^{(p)}(\rho,\nu_0)f_p(\nu)f_{p'}(\nu')
\delta_{pp'}\delta(\bmath{u}-\bmath{u}'),
\label{rawcov}
\end{equation}
where $\rho=|\bmath{u}|$ and $C^{(p)}(\rho,\nu_0)$ is the
ensemble-average power spectrum of the $p$th physical component at the
reference frequency $\nu_0$.  Since the sky is real, the Fourier modes
also obey $[a^{(p)}(\bmath{u},\nu)]^*=a^{(p)}(-\bmath{u},\nu)$.
We note that the assumption of rotational invariance means that the
ensemble-average power spectrum for each component is azimuthally
symmetric in the Fourier domain and hence a function of $\rho$.  For
$\ell \ga 60$, we make contact with the spherical harmonic expansion
by making the approximation
\[
C^{(p)}(\rho,\nu_0) \approx 
\left.c^{(p)}_\ell(\nu_0)\right|_{\ell = 2\pi\rho},
\]
which is accurate to within one per cent.  If we define the quantity
$D^{(p)}(\rho,\nu_0)=\rho^2 C^{(p)}(\rho,\nu_0)$, then in a similar
way we find
\[
D^{(p)}(\rho,\nu_0) \approx 
\frac{1}{(2\pi)^2}\left.d^{(p)}_\ell(\nu_0)\right|_{\ell = 2\pi\rho}.
\]

\subsection{Visibility data}
\label{visdata}

Expressing the temperature fluctuations of each component in terms of
the Fourier modes $a^{(p)}(\bmath{u}, \nu)$ is particularly useful
when discussing interferometer observations.
Assuming a small field size, and ignoring instrumental noise for the
moment, an interferometer measures samples from the complex {\em
  visibility plane} ${\cal S}(\bmath{u},\nu)$ arising from the sky
signal.  After converting to temperature units, this is given by
\begin{equation}
{\cal S}(\bmath{u},\nu) 
= \int A(\bmath{x},\nu) \frac{\Delta T(\bmath{x},\nu)}{T}
\exp(2\upi i\bmath{u}\cdot\bmath{x})~{\rm d}^2\bmath{x},
\label{visdef}
\end{equation}
where $\bmath{x}$ is the position relative to the phase centre,
$A(\bmath{x},\nu)$ is the (power) primary beam of the antennas at the
observing frequency $\nu$ (normalised to unity at its peak), and
$\bmath{u}$ is a baseline vector in units of wavelength.  
In the following discussion, we assume the primary beam
of a single antenna is circular symmetric and so does not vary with
parallactic angle. 

From (\ref{visdef}), we see that $S(\bmath{u},\nu)$ is the Fourier
transform of the product of the sky temperature fluctuations and the
primary beam, which is equivalent to the convolution of the underlying
Fourier modes $a(\bmath{u},\nu)$ with the Fourier transform of the
primary beam $\widetilde{A}(\bmath{u},\nu)$,
\begin{equation}
{\cal S}(\bmath{u},\nu) =
a(\bmath{u},\nu)\star\widetilde{A}(\bmath{u},\nu).
\label{eqn:aconvol}
\end{equation}
The positions in the $uv$-plane at which this function is sampled by
the interferometer are determined by the physical positions of its
antennas and the direction of the field on the sky. The samples
$\bmath{u}_j$ lie on a series of curves (or
$uv$-tracks), which we may denote by the function
$\widetilde{B}(\bmath{u},\nu)$ that equals unity where the Fourier
domain (or $uv$-plane) is sampled and equals zero elsewhere. The
function $\widetilde{B}(\bmath{u},\nu)$ may be inverse Fourier
transformed to give the {\em synthesised beam} $B(\bmath{x},\nu)$ of
the interferometer at an observing frequency $\nu$. For a realistic
interferometer, the sample values will also contain a contribution due
to instrumental noise. Thus, at an observing frequency $\nu$, the
$j$th baseline $\bmath{u}_j$ of an interferometer (measured in
wavelengths) measures the complex {\em visibility}
\[
{\cal V}(\bmath{u}_j,\nu) = {\cal S}(\bmath{u}_j,\nu) 
+ {\cal N}(\bmath{u}_j,\nu),
\]
where ${\cal N}(\bmath{u}_j,\nu)$ is the instrumental noise on the
$j$th visibility. 

\begin{figure}
\begin{center}
\includegraphics[width=6.7cm]{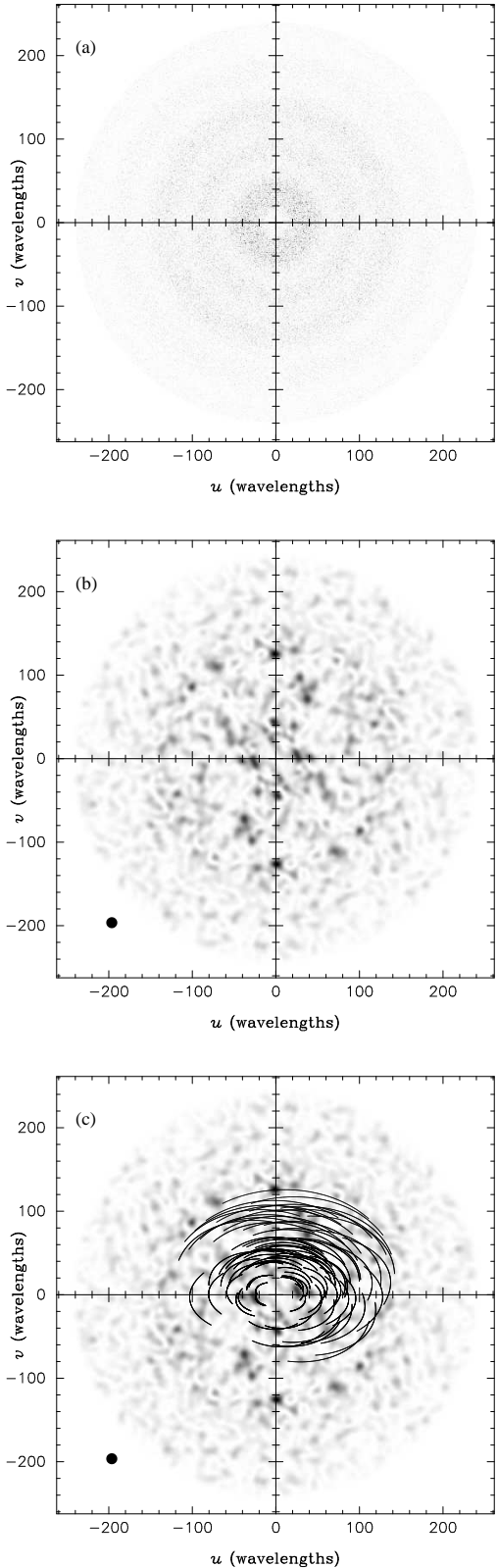}
\caption{(a) A realisation of the two-dimensional CMB power spectrum
  for an inflationary CDM model (see text). (b) The same realisation
  as in (a), but convolved with the aperture function of our model
  interferometer, assuming a Gaussian primary beam of 4.6$\degr$ FWHM.
  (c) The $uv$-coverage for a simulated 5-h observation with a
  14-element interferometer; each point corresponds to
  a sample with an integration time of 64-s.}
\label{fig1}
\end{center}
\end{figure}

The effects of the convolution of the underlying Fourier modes with
the aperture function and subsequent sampling are illustrated in
Fig.~\ref{fig1}, for the case when only CMB emission is present (and
neglecting instrumental noise).
In panel (a), we plot the modulus-squared of the underlying Fourier
modes for a realisation of an inflationary CDM model with a Hubble
parameter $H_0 =$~65~km~s$^{-1}$ Mpc$^{-1}$ and a fractional baryon
density $\Omega_{\rm b} = 0.05$. The model is spatially flat with
$\Omega_{\rm m}=0.3$ and $\Omega_\Lambda=0.7$, a primordial scalar
spectral index $n=1$, and no tensor modes.  The spectrum is normalised
to COBE.  We note the presence of acoustic `rings' in this
two-dimensional power spectrum.  We also see that the different
Fourier modes are uncorrelated.  In panel (b), we plot the same
realisation but after convolution with an aperture function
corresponding to an interferometer with a Gaussian primary beam of
4.6$\degr$ FWHM.  We see that the Fourier modes are no longer all
independent, but appear correlated on lengths scales corresponding to
the width of aperture function. The $1/e$ diameter of this function is
plotted as the solid disc bottom left-hand corner of the figure. In
panel (c) we overlay a typical 5-h $uv$-coverage 
for a compact VSA
observation of a single field at a declination of
30$\degr$. This instrument consists of 14 antennas with physical separations
lying between 20~cm and 1.50~m (see section~\ref{application}).
Each point corresponds to a sample with an integration time
of 64 s.

\subsection{Mosaicing}
\label{sec:mosaicing}

From~(\ref{eqn:aconvol}), we see that in order to obtain a high
resolution in Fourier space and thus in $\ell$-space, the Fourier
transform of the primary beam needs to be narrow, implying large
primary beams and small antennas. On the other hand, high sensitivity
demands a large collecting area for each antenna. Clearly, some
compromise has to be established between sensitivity and
$\ell$-resolution, and the current generation of CMB interferometers
differ in their respective priorities. However, the problem can be
circumvented by a technique called mosaicing (Ekers \& Rots 1979,
Cornwell 1988, Cornwell et al. 1993, Holdaway 1999). Mosaicing
consists of multiple observations of adjacent, overlapping fields on
the sky. This increases the effective beam size of the telescope and
thus the $\ell$-resolution. 

Mosaicing exploits that a telescope samples along each baseline a
whole superposition of visibilities. This is due to the finite extent
of the the antenna aperture, which means that effectively all
baselines corresponding to any two points on the two antennas that
make up the baseline are sampled. Observations with different pointing
centres probe different superpositions and can thus help to
disentangle the individual contributions (Ekers \& Rots 1979, Cornwell
1988). 

Unfortunately, observations of overlapping fields also require that we
take into account the correlations between the fields.  In
Appendix~\ref{sec:appndx:mosaicing} we show that the effect of
mosaicing may be be included by defining an {\em effective} primary
beam $A_{\rm eff}(\bmath{x},\nu)$ that replaces $A(\bmath{x},\nu)$ in
(\ref{visdef}).

\subsection{Non-coplanar baselines and $w$-corrections}
\label{sec:wcorrections}

For telescopes with a large primary beam $A(\bmath{x}, \nu)$, the
two-dimensional approximation~(\ref{visdef}) becomes inaccurate.
Radiation incident away from the telescopes pointing centres suffers
from phase errors if the $w$-components of the telescope baselines
(i.e. the component in the direction of the source) do not vanish, or
if the baselines are not coplanar. This is a geometric effect and is
usually referred to as $w$-distortion or as non-coplanar baselines
(see e.g. Cornwell \& Perley 1993; Perley 1999\nocite{perley}).

In order to obtain a good resolution in Fourier space, the VSA has a
comparatively large primary beam.  While the antennas are all mounted
on the same steerable table, they still track the source individually.
This delay--tracking helps to identify foregrounds or other systematic
errors by their different fringe rates. However, in the course of a
longer observation, the projection of the source onto the telescope
table changes and the array is effectively no longer coplanar.

This topic is discussed in detail in
Appendix~\ref{sec:appndx:wcorrections}, where we show that, similarly
to mosaicing, again the effects of the non--coplanar baselines may be
included by defining an appropriate {\em effective} primary beam
$A_{\rm eff}(\bmath{x},\nu)$ that replaces $A(\bmath{x},\nu)$ in
(\ref{visdef}). Thus, the following discussion can be
straightforwardly generalised to large fields.

\section{Preliminary analysis}
\label{prelimanal}

As discussed in HLJ95, before one can perform any meaningful
statistical analysis, it is necessary to remove as many unwanted
contributions as possible from the visibilities. It is also useful to
compress the data sufficiently, such that the computational burden
of the subsequent likelihood analysis is somewhat reduced.

\subsection{Removal of bad data and point sources}
\label{sec:baddatapointsources}

Preliminary analysis of the time-ordered visibility data consists of
the removal of periods of bad data due to poor weather conditions and
system failures, and the subtraction of the contribution to the sky
signal from identified discrete radio sources above some confusion
limit.  For example, O'Sullivan et al. (1995) discuss how this
preliminary data analysis was carried out for CMB observations with
the CAT interferometer, using the Ryle Telescope (Jones 1990) to
perform the radio source subtraction.  For VSA observations, the
systematic removal of periods of bad data are performed in an
analogous manner to that used for the CAT. The
subtraction of point sources from VSA data again relies on Ryle
Telescope observations to identify and determine the fluxes of all
sources above some flux limit in the observed field. This process is
also assisted by the use of two source-subtraction antennas situated
adjacent to the main VSA instrument. The details of the point source
identification and subtraction techniques used for VSA observations is
discussed in Taylor (2000) and Taylor et al. (2001).

After periods of bad data and point source emission have been removed,
the remaining non-cosmological contributions to the visibilities will
then be from unsubtracted point sources below the flux limit of the
observations, and from fluctuations in the Galactic free-free,
synchrotron and dust emission. These foregrounds are usually
identified by their spectral differences from the CMB, using
multifrequency observations.

\subsection{`Map'-making in the Fourier domain}
\label{sec:binning}

For the new generation of CMB interferometers, the total number of
visibilities at each frequency is very large. For example, the total
number of 64-s samples in a typical 5-h observation with a 14-element
interferometer is approximately 25,000, and any given field is 
observed for at least several days. It is therefore necessary to
compress these data in some way, as discussed in HLJ95. This is
analogous to the `map-making' step in the analysis of single-dish CMB
experiments, in which time-ordered data is binned into pixels on the
sky (see, for example, Borrill 1999). For an interferometer, however,
the visibilities at each observing frequency are binned into cells in
the $uv$-plane.

Since we are not interested in making accurate CMB maps from the
binned data, but are more concerned with estimating the CMB power
spectrum correctly, the $uv$-plane is simply divided into 
equal-area square cells (or pixels) of side $\Delta u$, within each of
which ${\cal S}(\bmath{u},\nu)$ is assumed to be constant. Clearly,
only a small number $N_c$ of these cells will contain observed 
visibility samples. We assemble
the (complex) values in each observed pixel into the (complex) vector
$\bmath{s}$ of length $N_c$.  We then calculate the maximum-likelihood
solution for these $N_c$ values as follows.

Suppose that the total number of visibility samples at the observing
frequency $\nu$ is $N_v$, and that the $j$th visibility is measured at
a point $\bmath{u}_j$ in the $uv$-plane. Let us assemble these
visibilities into the (complex) data vector $\mathbfss{v}$ of length
$N_v$, and also define the $N_v \times N_c$ pointing matrix with
elements
\[
M_{jk} = 
\left\{
\begin{tabular}{ll}
1 & \mbox{if $\bmath{u}_j$ lies in the $k$th cell}, \\
0 & \mbox{otherwise}.
\end{tabular}
\right.
\]
In this way, we may write the data vector as
\begin{equation}
\mathbfss{v} = \mathbfss{M}\bmath{s}+\mathbfss{n},
\label{eqn:mm1}
\end{equation}
where $\mathbfss{n}$ is the (complex) noise vector of length $N_v$,
whose $j$th element is simply ${\cal N}(\bmath{u}_j,\nu)$.  Assuming
that the instrumental noise on the real and imaginary parts of the
visibilities are independent and Gaussian, it is described by the
complex multivariate Gaussian probability distribution (see Eaton
1983)
\begin{equation}
\Pr({\mathbfss n}) = \frac{1}{\pi^{N_v} |{\mathbfss N}|}
\exp(-{\mathbfss n}^\dagger
{\mathbfss N}^{-1}{\mathbfss n}),
\label{eqn:mm2}
\end{equation}
where ${\mathbfss N}=\langle{\mathbfss nn}^\dagger\rangle$ is the 
instrumental noise covariance matrix. Substituting (\ref{eqn:mm1}) into
(\ref{eqn:mm2}), we find that the likelihood of obtaining the observed
visibility data given a particular signal vector $\bmath{s}$
is given by
\[
\Pr({\mathbfss v}|\bmath{s}) = 
\frac{1}{\pi^{N_v} |{\mathbfss N}|}
\exp[-({\mathbfss v}-{\mathbfss M}\bmath{s})^\dagger
{\mathbfss N}^{-1}({\mathbfss v}-{\mathbfss M}\bmath{s})].
\]
Maximising over $\bmath{s}$ then yields the maximum-likelihood
solution, which we will denote by $\bmath{v}$ (instead of the more usual
$\hat{\bmath{s}}$).  We find that this vector of {\em binned}
visibility data (of length $N_c$) is given in terms of the original
visibility data vector $\mathbfss{v}$ (of length $N_v$) by
\begin{equation}
\bmath{v} = ({\mathbfss M}^{\rm t}{\mathbfss N}^{-1}{\mathbfss M})^{-1}
{\mathbfss M}^{\rm t}{\mathbfss N}^{-1}\mathbfss{v}.
\label{eqn:mm3}
\end{equation}
Moreover, if we substitute (\ref{eqn:mm1}) back 
into (\ref{eqn:mm3}), we find that the
vector of binned visibilities can be written as
\[
\bmath{v} = ({\mathbfss M}^{\rm t}{\mathbfss N}^{-1}{\mathbfss M})^{-1}
{\mathbfss M}^{\rm t}{\mathbfss
N}^{-1}(\mathbfss{M}\bmath{s}+{\mathbfss n}) = \bmath{s}+\bmath{n},
\]
where 
\[
\bmath{n} = ({\mathbfss M}^{\rm t}{\mathbfss N}^{-1}{\mathbfss M})^{-1}
{\mathbfss M}^{\rm t}{\mathbfss N}^{-1}{\mathbfss n}.
\]
This shows that the binned
visibility data $\bmath{v}$ is the sum of the pixelised signal
$\bmath{s}$ and 
some residual pixelised noise $\bmath{n}$ with
covariance matrix
\begin{equation}
\bmath{N} = \langle\bmath{nn}^\dagger\rangle = 
({\mathbfss M}^{\rm t}{\mathbfss N}^{-1}{\mathbfss M})^{-1}.
\label{eqn:mm4}
\end{equation}

It is instructive to consider the special case in which the original
instrumental noise covariance matrix $\mathbfss{N}$ is diagonal, so
that the instrumental noise on each original visibility is uncorrelated.
In fact, this is usually the case for interferometer data, and we may write
${\mathbfss N}= \mbox{diag}(\sigma_1^2,\ldots,\sigma_{N_v}^2)$,
where $\sigma_j^2$ is the variance of the instrumental noise on the
$j$th original visibility. In this case, it is straightforward to show
that the maximum-likelihood solution (\ref{eqn:mm3}) reduces
to simple cell-averaging. Thus, the value $v_k$ of the binned
visibility in the $k$th cell is given by
\[
v_k = \frac{\sum_{j=1}^m{\cal V}_j/\sigma_j^2}{\sum_{j=1}^m 1/\sigma_j^2},
\]
where $m$ is the number of original visibilities lying in that cell.
Similarly, from (\ref{eqn:mm4}), we find that the residual pixelisation noise
has the covariance matrix $\bmath{N} = \mbox{diag}(\epsilon_1^2,\ldots,
\epsilon_{N_c}^2)$, where
\[
\epsilon_k^2 = \frac{1}{\sum_{j=1}^m 1/\sigma_j^2}.
\]

At each observing frequency $\nu$, it is the binned visibility
vector $\bmath{v}$ and pixelised noise vector $\bmath{n}$ that
constitute the basic data which are analysed in the estimation of the
CMB power spectrum. It is usual to associate the $k$th binned visibility
$v_k$ and noise $n_k$ with the the position $\bmath{u}_k$ that
corresponds to the centre of the $k$th cell in the $uv$-plane.
This is, however, not necessary. Indeed, one may preserve more
information regarding the distribution of samples in the
$uv$-plane by instead choosing $\bmath{u}_k$ to be the  
`centre-of-mass' of the positions of the visibility samples within the
$k$th cell. 

It only remains to choose the size of the cells $\Delta u$ in the
$uv$-plane. Since the speed with which the likelihood function can be
calculated is strongly dependent on the number of binned visibility
points that are analysed, we must make a compromise between 
how accurately the binned
data represent the original visibilities, and the desire to use fewer
data points in order to speed up the calculation. In fact, 
two natural limits exist for $\Delta u$. 

A minimum size for the $uv$ cells is given by the requirement that each
binned point on the grid represents an independent estimate of the
underlying visibility. If the typical sample time per visibility is
$\tau$ sec, then for samples
at a distance from the $uv$-origin of $|\bmath{u}|$, time-smearing
will correlate visibilities separated along a $uv$-track by less than
approximately
\[
\Delta u \approx 2\pi|\bmath{u}| \frac{\tau}{24\times 60\times 60}.
\]
For $\tau = 64$ sec and $uv$-coverage which extends out to
$|\bmath{u}| \approx 200$ wavelengths, then $\Delta u \ga 1$ wavelengths.

A maximum size for the $uv$ cells is derived by considering the fact
the measured visibilities are obtained by a convolution of the
underlying Fourier modes with the interferometer aperture function.
For example, the VSA in its compact configuration, this function
is a Gaussian a FWHM of approximately 12 wavelength. Hence, as
discussed by HLJ95, from the sampling theorem we require
$\Delta u \la 6$ wavelengths. Nevertheless, finer binning is 
still useful since it provides better
approximation for the position of the visibilities in the $uv$-plane.
Thus, we find that the $uv$ cell size is reasonably well determined by
our two limiting cases and for the simulations considered here we
use $\Delta u = 3$ wavelengths. 

\begin{figure}
\begin{center}
\includegraphics[angle=-90,width=7cm]{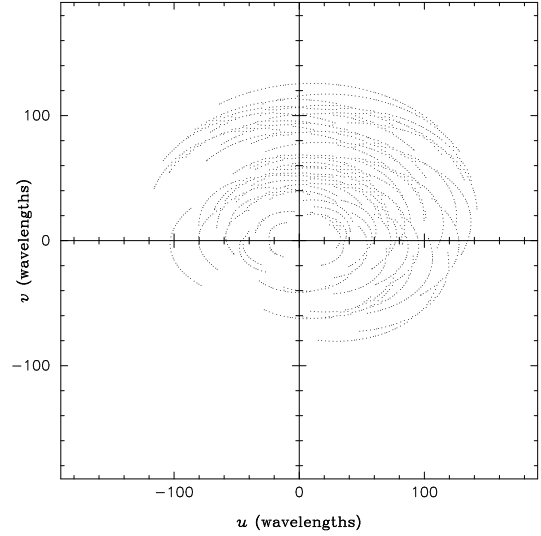}
\caption{An illustration of the positions of the 
binned visibilities resulting from the
samples shown in Fig.~\ref{fig1}(c) using a cell size $\Delta u = 3$
wavelengths.}
\label{fig2}
\end{center}
\end{figure}

As an illustration, in Fig.~\ref{fig2} we plot the positions of the
binned visibilities corresponding to the samples shown in 
Fig.~\ref{fig1}(c). 
Each dot shows the position
$\bmath{u}_k$ of the `centre-of-mass' of each cell, with which the binned
visibility $v_k$ is associated.
In this case, the number
of non-zero binned visibilities is approximately 2500.
Provided the geometry of the interferometer is not
varied, each period of observation of a given field will produce
visibility samples lying in the same set of cells in the $uv$-plane.
Thus, after binning the {\em total} number of complex data points at each
observing frequency that are used to estimate the CMB power spectrum 
is about 2500.

\section{The visibilities covariance matrix}
\label{viscovmat}

After binning, the data
consist of the binned visibility data vector $\bmath{v}(\nu_i)$ 
$(i=1,\ldots,N_f)$ at each of the $N_f$ observing frequencies, together
with the corresponding noise vector $\bmath{n}(\nu_i)$ at each
frequency. To keep our notation simple, we combine {\em all} the
binned visibilities into the {\em single} data vector $\bmath{v}$, and
all the noise vectors into a single vector $\bmath{n}$. 

Let us assume that the $j$th
binned visibility is measured at a point $\bmath{u}_j$ in the $uv$-plane and
at an observing frequency $\nu_j$. From this point onwards, we shall 
denote the $j$th {\em binned} visibility by ${\cal V}_j$. 
Similarly, we shall denote the
noise on the $j$th binned visibility by ${\cal N}_j$ and the
contribution
from the sky signal by ${\cal S}_j$.
With a view to a practical 
implementation, it is easier to pursue the analysis in terms
of the real and imaginary parts of each visibility, which we write as
\[
{\cal V}_j \equiv {\cal V}_j^{(R)} + i {\cal V}_j^{(I)}.
\]
We may also divide the signal and noise contributions
${\cal S}$ and ${\cal N}$ into their real and imaginary parts in the
same way. Thus, if the interferometer observation (including all
observing frequencies) consists of total number 
$N_t$ of complex binned visibilities, we adopt the convention that
the data vector $\bmath{v}$ is of length $N_d=2N_t$, consisting
of first the real parts and then the imaginary parts of the
visibilities, i.e.
\[
v_j = {\cal V}_j^{\rm (R)},\qquad v_{N_t+j} = {\cal V}_j^{\rm (I)}.
\]
It is also useful to write $\bmath{v}=\bmath{s}+\bmath{n}$, where
the signal vector $\bmath{s}$ 
and noise vector $\bmath{n}$ are defined in a similar way.

Assuming there are no correlations between the contributions to
visibilities from the sky signal and the instrumental noise,
the covariance matrix of the interferometer data is given by
\[
\bmath{C} = \langle \bmath{v}\bmath{v}^{\rm t} \rangle
=\langle \bmath{s}\bmath{s}^{\rm t} \rangle
+\langle \bmath{n}\bmath{n}^{\rm t} \rangle 
=\bmath{S} + \bmath{N},
\]
where $\bmath{S}$ and $\bmath{N}$ are respectively the
covariance matrices of the contributions from the
sky signal and noise. To an excellent approximation, 
$\bmath{N}$ make be taken as diagonal, whereas the signal covariance
matrix $\bmath{S}$ has the structure
\begin{equation}
\bmath{S} = \left(
\begin{array}{l|l}
\bmath{S}^{(R,R)} & \bmath{S}^{(R,I)} \\ \hline
\bmath{S}^{(I,R)} & \bmath{S}^{(I,I)} \\
\end{array}
\right),
\label{eqn:smatdef}
\end{equation}
where $S_{ij}^{(R,R)} = \langle {\cal S}^{(R)}_i {\cal S}^{(R)}_j \rangle$
etc. 

As shown in Appendix~\ref{sec:appndx:covariance}, for an observation
of a single field in the flat-sky approximation, the elements of
$\bmath{S}$ may be conveniently written in terms of the complex
correlators
\begin{eqnarray*}
\langle {\cal S}_i {\cal S}_j^\ast \rangle \!\!\! &=&\!\!\!
\int\!\!\!\int\! \tilde{A} (\bmath{u}_i - \bmath{u}, \nu_i)
\tilde{A}^\ast (\bmath{u}_j - \bmath{u}, \nu_j) 
\xi(|\bmath{u}|,\nu_i,\nu_j)~{\rm d}^2\bmath{u}, \\
\langle {\cal S}_i {\cal S}_j\rangle\!\!\! &=& \!\!\!
\int\!\!\!\int\! \tilde{A} (\bmath{u}_i - \bmath{u}, \nu_i)
\tilde{A} (\bmath{u}_j + \bmath{u}, \nu_j) 
\xi(|\bmath{u}|,\nu_i,\nu_j)~{\rm d}^2\bmath{u},
\end{eqnarray*}
where $\tilde{A}(\bmath{u},\nu)$ is the aperture function of the
interferometer horns and $\xi(\rho,\nu_i,\nu_j)$
is the {\em generalised power spectrum}, which is defined by
\begin{equation}
\xi(\rho,\nu_i,\nu_j)=
\sum_{p=1}^{N_p} f_p(\nu_i)f_p(\nu_j)C^{(p)}(\rho,\nu_0).
\label{eqn:defgenps}
\end{equation}
We note that, if the primary beam of the interferometer horns is
is symmetric with respect to inversion through the origin,
i.e. $A(\bmath{x},\nu) = A(-\bmath{x},\nu)$, then the aperture
function $\tilde{A}(\bmath{x},\nu)$ is real. In this case, 
$\bmath{S}^{(R,I)}$ and $\bmath{S}^{(I,R)}$ vanish, and so
$\bmath{S}$ in (\ref{eqn:smatdef}) is block-diagonal.

By introducing
polar coordinates $(\rho,\theta)$ into the $uv$-plane, 
such that $\bmath{u} = \rho (\cos\theta,\sin\theta)^{\rm t}$,
we can also write $\langle {\cal S}_i {\cal S}_j^\ast \rangle$
and $\langle {\cal S}_i {\cal S}_j\rangle$ formally as the one-dimensional
integrals
\begin{eqnarray}
\langle {\cal S}_i {\cal S}_j^\ast \rangle &=&
\int_0^\infty {\rm d}\rho\,\rho \,W^{(1)}_{ij}(\rho)
\,\xi(\rho,\nu_i,\nu_j), \label{eqn:ss*}\\
\langle {\cal S}_i {\cal S}_j\rangle &=&
\int_0^\infty {\rm d}\rho\,\rho \,W^{(2)}_{ij}(\rho)
\,\xi(\rho,\nu_i,\nu_j),\label{eqn:ss}
\end{eqnarray}
where the {\em window functions} $W^{(1)}_{ij}(\rho)$ and
$W^{(2)}_{ij}(\rho)$ are defined respectively by
\begin{eqnarray*}
W^{(1)}_{ij}(\rho)
& = & \int_0^{2\pi} {\rm d}\theta\, 
\tilde{A}(\bmath{u}_i-\bmath{u},\nu_i)
\tilde{A}^\ast (\bmath{u}_j-\bmath{u},\nu_j), \\
W^{(2)}_{ij}(\rho)
& = & \int_0^{2\pi} {\rm d}\theta\, 
\tilde{A}(\bmath{u}_i-\bmath{u},\nu_i)
\tilde{A}(\bmath{u}_j+\bmath{u},\nu_j).
\end{eqnarray*}
The advantage of writing $\langle {\cal S}_i {\cal S}_j^\ast \rangle$
and $\langle {\cal S}_i {\cal S}_j \rangle$ in this way is that it
provides a natural separation between the influence on $\bmath{S}$
of instrumental effects and the underlying power spectrum of the sky.
In particular, for an interferometer with a given geometry, 
the window functions $W^{(1)}_{ij}(\rho)$ and
$W^{(1)}_{ij}(\rho)$ are fixed, independent of the underlying
sky power spectrum. Thus, for a given array geometry, 
these quantities need only be calculated once.

As discussed in Appendices~\ref{sec:appndx:mosaicing} and
\ref{sec:appndx:wcorrections}, if one performs mosaiced observations
or includes the effects of non-coplanar baselines, the correlators $\langle
{\cal S}_i {\cal S}_j^\ast \rangle$ and $\langle {\cal S}_i {\cal
  S}_j^\ast \rangle$ may still be written in the above forms, 
but with $\tilde{A}(\bmath{u},\nu)$ replaced by an
appropriate {\em effective} aperture function $\tilde{A}_{\rm
  eff}(\bmath{u},\nu)$. The relevant expressions for $\tilde{A}_{\rm
  eff}$ are given in (\ref{eqn:effapermos2}) 
and (\ref{eqn:effaperw}). In general, $\tilde{A}_{\rm
  eff}$ is not a real function, and so the signal covariance matrix 
(\ref{eqn:smatdef}) is not, in general, block-diagonal.

\subsection{Sparse structure of the covariance matrix}
\label{sparsestructure}

The expressions (\ref{eqn:ss*}) and (\ref{eqn:ss}) 
provide general formulae for
calculating the element $S_{ij}$ of the signal covariance matrix.  So
far we have assumed that all these elements  must be calculated.
If, for example, the number of binned complex visibilities is about $2500$, 
the real data vector $\bmath{v}$ has about $5000$ elements.
Thus, in general, the construction of the (symmetric) covariance matrix
$\bmath{S}$ requires the evaluation of approximately 
$(5000\times 5000)/2$ elements for
any given generalised power spectrum $\xi(\rho,\nu_i,\nu_j)$.
It is, however, straightforward to show that the
signal covariance matrix $\bmath{S}$ is in fact
very sparse and hence the computational requirements 
are somewhat reduced. 


%
\begin{figure}
\begin{center}
\includegraphics[width=6.5cm]{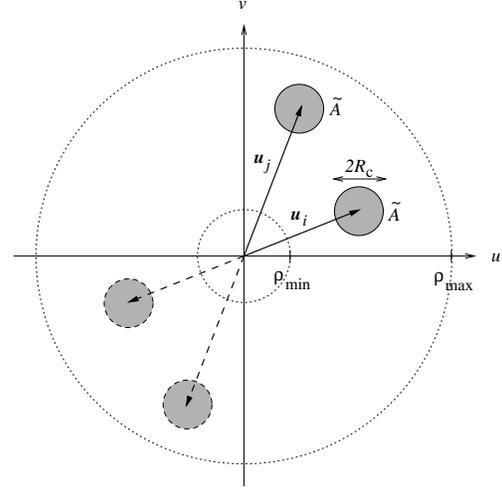}
\caption{An illustration of the sparse nature of the signal 
  covariance matrix $\bmath{S}$ for interferometer observations.}
\label{fig3}
\end{center}
\end{figure}

The reason for the sparse nature of the signal covariance matrix is
illustrated in Fig.~\ref{fig3}, and is peculiar to observations of the
CMB made with interferometers. For simplicity let us consider a
single-field observation with no non-coplanar baselines, and focus
particularly on two measured visibilities at positions $\bmath{u}_i$
and $\bmath{u}_j$ respectively in the $uv$-plane.  From~(\ref{eqn:aconvol}),
these visibilities are obtained simply by convolving the underlying
independent Fourier modes by the aperture function of the
interferometer, which is illustrated by the solid shaded circles in the
figure.

Since the extent of the antenna function is determined simply by the
physical size of the interferometer horns, it is identically zero
outside some critical radius $R_c$.  Hence one would expect that, if
the separation $|\bmath{u}_i-\bmath{u}_j|$ of the two visibility
samples were greater than $2R_c$, then the corresponding element
$S_{ij}$ of the signal covariance matrix should be {\em identically zero}.
An additional subtlety remains, however, because the sky is real, and
so $a(-\bmath{u},\nu) = a^*(\bmath{u},\nu)$.  Consequently,
correlations exist between areas of the Fourier domain that lie on
diametrically opposite sides of the origin of the $uv$-plane; this is
illustrated in Fig.~\ref{fig3}. Thus, the element $S_{ij}$ of the
signal covariance matrix will be identically zero provided
$\bmath{u}_i$ and $\bmath{u}_j$ satisfy {\em both} the conditions
\begin{eqnarray*}
|\bmath{u}_i-\bmath{u}_j| & > & 2R_c, \\
|\bmath{u}_i+\bmath{u}_j| & > & 2R_c.
\end{eqnarray*}
With reference to Fig.~\ref{fig3}, this occurs when no two shaded
discs overlap.

We may obtain a simple estimate for the sparsity fraction of the
matrix $\bmath{S}$ as follows. Suppose the interferometer samples the
$uv$-plane with roughly uniform two-dimensional coverage in the range
$\rho_{\rm min}$ to $\rho_{\rm max}$. Thus, given the position of the
sample $\bmath{u}_i$, a second sample $\bmath{u}_j$ may lie anywhere
in region between the two concentric circles $\rho=\rho_{\rm min}$ and
$\rho=\rho_{\rm max}$ shown in Fig.~\ref{fig3}, which is of area
$\pi(\rho_{\rm max}^2-\rho_{\rm min}^2)$. In order for $S_{ij}$ to be
non-zero, however, $\bmath{u}_j$ must lie within a distance $2R_c$ of
the point $\bmath{u}_i$ or $-\bmath{u}_i$, which corresponds to an
available area $2\pi(2R_c)^2$ of the $uv$-plane.  Thus, an estimate of
the sparsity fraction $f_s$ is given by the ratio of the areas, which
reads
\[
f_s = \frac{8R_c^2}{\rho_{\rm max}^2-\rho_{\rm min}^2}.
\]
For example, the VSA in compact configuration has $R_c \approx 8$,
$\rho_1 \approx 20$ and $\rho_2 \approx 150$ wavelengths and so the
sparsity fraction in this case is $f_s \approx 0.02$, i.e. only about
2 per cent of the elements are non-zero.

In fact, the fraction of elements $S_{ij}$ that need be calculated in
practice is somewhat smaller than $f_s$. As stated above, the element
$S_{ij}$ is identically zero if $\bmath{u}_i$ and $\bmath{u}_j$
satisfy the constraints, with $R_c$ equalling the full extent of the
aperture function. For most interferometers, however, the aperture
function falls close to zero some way before $R_c$. Thus, to a very
good approximation, $S_{ij} \approx 0$ if
\begin{eqnarray}
|\bmath{u}_i-\bmath{u}_j| & > & 2\alpha R_c, \label{eqn:sp1}\\
|\bmath{u}_i+\bmath{u}_j| & > & 2\alpha R_c, \label{eqn:sp2}
\end{eqnarray}
where $\alpha$ may be taken to be somewhat less than unity. The resulting
sparsity fraction $f_s$ is then lowered by a factor $\alpha^2$. 

In particular, as we describe below, many interferometers have an
aperture function that are well described by a (truncated)
two-dimensional Gaussian.  In this case, we find empirically that an
excellent approximation to $\bmath{S}$ is obtained by assuming to be
zero all those elements $S_{ij}$ for which (\ref{eqn:sp1}) and 
(\ref{eqn:sp2}) are
satisfied, where $\alpha$ is chosen so that $\alpha R_c$ corresponds
to the radius at which the Gaussian aperture function drops to $1/e$
of its peak central value. This leaves only a very small percentage of
matrix elements that must be calculated by evaluating the integrals
(\ref{eqn:ss*}) and (\ref{eqn:ss}).  
On using this approximation for $\bmath{S}$ in
the subsequent maximum-likelihood estimation of the CMB power
spectrum, discussed later in section~\ref{application}, we find that the flat
band-power estimates obtained in each $\ell$-bin alter by less than 1
per cent, as compared to the case in which we set $\alpha=1$.

\subsection{The flat band-power parameterisation}
\label{flatbps}

It has become standard practice to parameterise the CMB power spectrum
in terms of {\em flat band-powers} as follows.  The range of $\rho =
|\bmath{u}|$ over which we wish to estimate the power spectra of each
component is divided into $N_b$ separate bins. These bins correspond
to separate annuli in the $uv$-plane, centred on the origin. One then
assumes that the quantity
\[
D^{(p)}(\rho,\nu_0) \equiv  \rho^2 C^{(p)}(\rho,\nu_0),
\]
for each physical component $(p=1,2,\ldots,N_p)$ is constant within
any given spectral bin. We denote the constant value of
$D^{(p)}(\rho,\nu_0)$ in the $b$th bin by $\overline{D}_b^{(p)}$, and
these quantities are the parameters whose values are to be determined
from the likelihood analysis described in
section~\ref{sec:likeanalysis}. Clearly, the
total number of parameters is $N_bN_p$.

This parameterisation of the component power spectra has the advantage
that, once the elements of signal covariance matrix have been
calculated for one set of band-power values $\overline{D}_b^{(p)}$,
it may be evaluated for any other set at almost no
extra computational cost.
This is easily demonstrated by first introducing the top-hat
function $T(a,b)$, which equals unity in the range $a\le \rho < b$ and
is zero elsewhere. We may then write $D^{(p)}(\rho,\nu_0)$ in terms of
our model parameters $\overline{D}^{(p)}_b$ as
\[
D^{(p)}(\rho,\nu_0) 
= \sum_{b=1}^{N_b} \overline{D}_b^{(p)} T(\rho_b,\rho_{b+1}),
\]
where $\rho_b$ the lower limit of the $b$th bin. Therefore, the
generalised power spectrum $\xi(\rho,\nu_i,\nu_j)$ in
(\ref{eqn:defgenps}) 
is now
given by
\[
\xi(\rho,\nu_i,\nu_j)=
\frac{1}{\rho^2}
\sum_{p=1}^{N_p} f_p(\nu_i)f_p(\nu_j)
\sum_{b=1}^{N_b} \overline{D}_b^{(p)} T(\rho_b,\rho_{b+1}).
\]
Hence the expressions (\ref{eqn:ss*}) and (\ref{eqn:ss}) can be written as
\begin{eqnarray}
\langle {\cal S}_j{\cal S}_j^\ast\rangle
& = & \sum_{p=1}^{N_p} f_p(\nu_i)f_p(\nu_j)
\sum_{b=1}^{N_b} \overline{D}_b^{(p)}
K^{(1)}_{ij}(\rho_b,\rho_{b+1}),\label{eqn:ss*2}\\
\langle {\cal S}_i{\cal S}_j\rangle
& = & \sum_{p=1}^{N_p} f_p(\nu_i)f_p(\nu_j)
\sum_{b=1}^{N_b} \overline{D}_b^{(p)}
K^{(2)}_{ij}(\rho_b,\rho_{b+1}),\label{eqn:ss2}\\
\end{eqnarray}
where the sets of integrals $K^{(r)}_{ij}(\rho_b,\rho_{b+1})$ (for
$r=1,2$) are given by
\begin{equation}
K^{(r)}_{ij}(\rho_b,\rho_{b+1}) =
\int_{\rho_b}^{\rho_{b+1}} \frac{{\rm d}\rho}{\rho} W^{(r)}_{ij}(\rho).
\label{eqn:integrals}
\end{equation}

As stated earlier, for a given array geometry, the window
functions $W^{(r)}_{ij}(\rho)$ $(r=1,2)$ are fixed.  Thus the
integrals $K^{(r)}_{ij}(\rho_b,\rho_{b+1})$ need only be evaluated
once, at the start of the calculation.  Then for any values of the
parameters $\overline{D}_b^{(p)}$, we may use (\ref{eqn:ss*2}) 
and (\ref{eqn:ss2}) to
evaluate the elements of the signal covariance matrix $\bmath{S}$.

We may choose the width of the bins arbitrarily, but an obvious choice
is the characteristic width of the aperture function of the
interferometer (for example, the $1/e$ width). Since this width is the
typical correlation length in the convolved Fourier domain, the errors
on the derived values of the flat band-powers in different bins will
be quasi-uncorrelated. The compact VSA, for example, samples the range
$\rho \approx $ 20--150 wavelengths and the aperture function for a
single field observation has a $1/e$ width of about $15$ wavelengths.
Thus, the total number of spectral bins is $N_b \approx 10$.

\subsection{Gaussian primary beam}
\label{gausspb}

So far we have not assumed a specific form for the primary beam
$A(\bmath{x},\nu)$ of the interferometer horns. Nevertheless, for many
interferometers, the primary beam may be modelled to a good
approximation as a two-dimensional Gaussian. Thus, the primary
beam at the $i$th observing frequency $\nu_i$ by
\[
A(\bmath{x},\nu_i)=\exp(-|\bmath{x}|^2/2\sigma_i^2),
\]
where $\sigma_i$ is the dispersion of the primary beam at this frequency.
The aperture function is simply the Fourier transform of the primary
beam, and is given by
\[
\tilde{A}(\bmath{u},\nu_i)=2\pi\sigma_i^2
\exp(-2\pi^2\sigma_i^2|\bmath{u}|^2).
\]

As we show in Appendix~\ref{sec:appndx:gaussian}, this assumption
allows us to obtain analytical expressions for the window functions
$W^{(r)}_{ij}(\rho)$ $(r=1,2)$, even for mosaiced observations and
including non-coplanar baselines. Thus the integrals $K^{(r)}_{ij}(\rho)$ in
(\ref{eqn:integrals}) can be evaluated quickly and straightforwardly
using standard numerical quadrature techniques. For a compact VSA
observation of a single field, the calculation of the 
integrals corresponding to non-zero elements of the 
signal covariance matrix $\bmath{S}$ require about 2 mins of CPU
time.\footnote{Throughout this paper CPU times refer to computations
performed on a single Intel Pentium III 1 GHz processor.}

\section{Likelihood analysis}
\label{sec:likeanalysis}

As discussed in the start of section~\ref{sec:binning}, the basic data
consist of the (real) binned visibility data vector $\bmath{v}$ of
length $N_d$, and the associated (real) residuals covariance matrix
$\bmath{N} = \langle \bmath{n}\bmath{n}^{\rm t}\rangle$. Assuming that
the underlying Fourier modes of the sky and the instrumental noise are
both Gaussian-distributed with zero mean, the likelihood of obtaining
the data, given some particular set of model parameter values
$\bmath{a}$, is
\[
{\cal L}(\bmath{v}|\bmath{a}) =
\frac{1}{(2\pi)^{N_d/2}|\bmath{C}(\bmath{a})|^{1/2}}
\exp[-\tfrac{1}{2}\bmath{v}^{\rm t}\bmath{C}^{-1}(\bmath{a})\bmath{v}]
\]
where $\bmath{C}(\bmath{a})=\bmath{S}(\bmath{a})+\bmath{N}$ is the sum
of the signal and noise covariance matrices. In our case, the vector
of model parameters $\bmath{a}$ contains the $N_b \times N_p$ flat
band-powers $\overline{D}_b^{(p)}$ in each spectral bin for each
physical component of sky emission and the structure of the
corresponding signal covariance matrix $\bmath{S}(\bmath{a})$ is
discussed in the previous section.  In fact, in order to avoid
numerical instabilities, we adopt the standard technique of working
instead with the log-likelihood function
\begin{equation}
\ln{\cal L}(\bmath{v}|\bmath{a}) =\mbox{constant}-
\tfrac{1}{2}\left[\ln |\bmath{C}(\bmath{a})|
+\bmath{v}^{\rm t}\bmath{C}^{-1}(\bmath{a})\bmath{v}\right].
\label{likedef2}
\end{equation}
%

The aim of any likelihood analysis is to find the parameter values
$\hat{\bmath{a}}$ that maximise the log-likelihood function,
and also obtain an estimate of the accuracy with which these
parameters have been determined.  There exist several different
strategies for achieving these goals, which we now discuss, with
particular emphasis on the computational cost of each approach when
applied to interferometer data.

\subsection{Evaluation of the multi-dimensional likelihood}
\label{evalmulti}

Ideally, one would like to evaluate the full (log-)likelihood function
over some hypercube in the space of parameters $\bmath{a}$. In this
way, the location of the (global) maximum is obtained immediately, and
the presence of multiple subsidiary maxima is readily observed.  Also,
if one is interested only in (say) the CMB power spectrum, one can
integrate out (or marginalise) the parameters describing the power
spectrum of Galactic emission in a straightforward manner.  Moreover,
marginalised distributions for individual parameters can be calculated
trivially, in order to obtain confidence limits.

Unfortunately, evaluation of the full likelihood function on a
hypercube in parameter space is numerically unfeasible when the number
of parameters is large. In our case, the parameter space has $N_bN_p$
dimensions. Thus, if the likelihood function were evaluated at $M$
points along each parameter axis, the total number of evaluations of
the log-likelihood function (\ref{likedef2}) is $M^{N_bN_p}$.  Even if one
assumes that the CMB is the only physical component of emission (i.e.
$N_p=1$), the dimensionality of the parameter space is equal to the
number of spectral bins in which one wishes to estimate the CMB power
spectrum. For the latest generation of CMB interferometers $N_b
\approx 10$, and so the parameter space has a high dimensionality. If,
in this case, the likelihood function were calculated at only $M=10$
points on each `axis', this requires $10^{10}$ evaluations of the
function (\ref{likedef2}).

Nevertheless, we note in passing that, with the advent of faster
computers and efficient algorithms, it has recently become possible to
sample directly from a multidimensional likelihood function using
Markov-Chain Monte-Carlo (MCMC) techniques (see e.g.
Christensen et al. 2001). Even in a space of
large dimensionality, these algorithms allow one to construct accurate
one-dimensional marginalised distributions for each parameter, using
relatively few evaluations of the likelihood function (typically
$\sim\mbox{ few}\times 1000$). The application of MCMC techniques to
the estimation of the CMB power spectrum from interferometer data will
be presented in a forthcoming paper.

\subsubsection{Evaluation of the log-likelihood function}
\label{evallfcn}

Despite the difficulties associated with the direct calculation of the
full likelihood over a hypercube in parameter space, it is
necessary for our later discussion
to consider the computational task associated with evaluating the
log-likelihood function (\ref{likedef2}) for a given set of parameter values
$\bmath{a}$.

First, one must calculate the corresponding covariance matrix
$\bmath{C}(\bmath{a})=\bmath{S}(\bmath{a})+\bmath{N}$.  Since
$\bmath{N}$ is constant, one needs only to calculate the signal
covariance matrix $\bmath{S}(\bmath{a})$.  As discussed in
section~\ref{flatbps}, after evaluating the integrals
(\ref{eqn:integrals}) once, at the
start of the calculation, the signal covariance matrix
$\bmath{S}(\bmath{a})$, for a given set of parameter values
$\bmath{a}$, may be calculated very quickly using (\ref{eqn:ss*2}) and
(\ref{eqn:ss2}). Since, for most interferometers, $\bmath{N}$ is diagonal to
a very accurate approximation, $\bmath{C}(\bmath{a})$ inherits the
sparse structure of $\bmath{S}(\bmath{a})$ described in
section~\ref{sparsestructure}.

Once $\bmath{C}$ has been calculated, one proceeds to evaluate the
log-likelihood function (\ref{likedef2}), which requires the evaluation of the
quadratic form $\bmath{v}^{\rm t} \bmath{C}^{-1}\bmath{v}$ and the
determinant $|\bmath{C}|$.  The standard procedure is first to perform
the Cholesky decomposition
\begin{equation}
\bmath{C}=\bmath{LL}^{\rm t},
\label{eqn:cholesky}
\end{equation}
where $\bmath{L}$ is a
lower triangular matrix. 
The determinant $|\bmath{C}|$ is then given by
\[
|\bmath{C}| = 
|\bmath{L}\bmath{L}^{\rm t}|=|\bmath{L}|^2,
\]
where, since $\bmath{L}$ is lower triangular, $|\bmath{L}|$ is simply
the product of its diagonal elements. To evaluate the quadratic form,
one then solves for $\bmath{x}$ the lower triangular system
\begin{equation}
\bmath{Lx}=\bmath{v},
\label{eqn:ltriang}
\end{equation}
which is computationally straightforward. Finally, one forms the scalar product
of $\bmath{x}$ with itself to obtain
\[
\bmath{x}^{\rm t}\bmath{x}
= \bmath{v}^{\rm t} (\bmath{L}^{-1})^{\rm t} 
\bmath{L}^{-1} \bmath{v}
= \bmath{v}^{\rm t} (\bmath{L}\bmath{L}^{\rm t})^{-1} 
\bmath{v}
= \bmath{v}^{\rm t} \bmath{C}^{-1} \bmath{v}.
\]

By far the most computationally intensive step in the calculation is
the Cholesky decomposition (\ref{eqn:cholesky}). 
For a compact VSA observation of
a single field, for example, the (real) binned visibility data vector
$\bmath{v}$ contains about $N_d=5000$ elements, so that $\bmath{C}$ has
dimensions $5000 \times 5000$. Using a standard {\sc lapack}
subroutine (Anderson et al. 1999) 
to perform the decomposition requires about 5 mins
of CPU time. The subsequent
solution of the lower-triangular system (\ref{eqn:ltriang}) is much faster,
requiring only 0.25 secs of CPU time. Hence, the total time required to
evaluate the log-likelihood function (\ref{likedef2}) for a given set of
parameters $\bmath{a}$ is around 5 mins.  We therefore see that the
evaluation of the full likelihood function over some hypercube in
parameter space is computationally unfeasible.

We note, in passing, that the {\sc lapack} library uses the
highly-optimised Basic Linear Algebra Subroutines ({\sc blas}) for
simple operations. In practice, this combination provides the fastest
readily-available dense matrix computational subroutines. As a
comparison, performing the same calculation of the log-likelihood
using simple unoptimised routines, such as those presented in Press et
al. (1994), required approximately 20 times the CPU time on an
equivalent processor.

\subsubsection{Sparse matrix conjugate-gradient algorithm}
\label{sparsecg}

As stated earlier, the covariance matrix $\bmath{C}$ is very
sparse for interferometer data.  Therefore, using standard dense
matrix subroutines, such as those in the {\sc lapack} library, is
wasteful both computationally and in terms of memory usage. One should
instead employ sparse matrix algorithms, which take advantage of the
sparse structure of $\bmath{C}$ and require only the storage of its
non-zero elements (plus an integer array to store their positions in
the matrix).  It was found that the most computationally efficient
approach was to solve the linear system
\begin{equation}
\bmath{Cx}=\bmath{v},
\label{eqn:clinear}
\end{equation}
with a preconditioned conjugate-gradient algorithm that performed its
internal matrix and vector operations using sparse matrix routines.
In particular, the preconditioner was chosen to be the sparse
incomplete Cholesky decomposition of $\bmath{C}$.  This matrix has the
form
\begin{equation}
\tilde{\bmath{C}}=\bmath{L}\bmath{D}\bmath{L}^{\rm t},
\label{eqn:icholesky}
\end{equation}
where $\bmath{D}$ is a diagonal matrix, and $\bmath{L}$ is lower
triangular with unit diagonal elements and which has the same sparse
structure as the lower triangle of $\bmath{C}$ (which is symmetric).
Since one imposes this sparse structure on $\bmath{L}$, the matrix
$\tilde{\bmath{C}}$ is only an approximation to $\bmath{C}$ (see
Greenbaum 1997). Nevertheless, the approximation is sufficiently
accurate to provide an extremely effective preconditioner, which
allows the conjugate-gradient algorithm to converge to the solution
$\bmath{x}$ in a just a few iterations.  Moreover, it was found that
the preconditioner $\tilde{\bmath{C}}$ was a sufficiently accurate
approximation to $\bmath{C}$ that the determinant
\[
|\tilde{\bmath{C}}| = |\bmath{L}\bmath{D}\bmath{L}^{\rm t}|
=|\bmath{L}||\bmath{D}||\bmath{L}^{\rm t}|
=|\bmath{D}|,
\]
differed by less than 0.01 per cent from the true determinant
$|\bmath{C}|$. Since $\bmath{D}$ is diagonal, its determinant is
trivially evaluated. Thus, the sparse matrix preconditioned
conjugate-gradient algorithm provides an efficient method of calculating the
complete log-likelihood function (\ref{likedef2}).

As in the dense matrix approach, the most computationally demanding
step of the calculation is the construction of the incomplete Cholesky
preconditioner (\ref{eqn:icholesky}).  As an example, for the $5000\times 5000$
covariance corresponding to a compact VSA observation of a single
field, the calculation requires about 30 sec of CPU time.
The subsequent preconditioned conjugate-gradient
solution of the linear system (\ref{eqn:clinear}) 
then requires only about 0.5 sec
of CPU time. Thus, we see that, in this case, 
the evaluation of the log-likelihood
function by this method is approximately 10 times faster than the
standard dense matrix approach using {\sc lapack} subroutines.
For larger problems, the {\sc lapack} routines scale as $O(N_d^3)$,
where $N_d$ is the size of the data vector, whereas
the sparse matrix algorithm scales as $O(f_s^{3/2}N_d^3)$, where
$f_s$ is the sparsity fraction of the covariance matrix $\bmath{C}$.
Unfortunately, the speed-up provided by the sparse matrix approach
is still insufficient to enable the
calculation of the full likelihood distribution over some hypercube in
parameter space. 

\subsection{Numerical maximisation of the likelihood}
\label{nummax}

When it is unfeasible to calculate the full likelihood function
directly over the parameter space, one usually resorts to numerical
maximisation.  The main disadvantage of this approach is that most
standard methods will converge to the nearest {\em local} maximum, which
need not be the global maximum. One would hope, however, that the
likelihood function is sufficiently structureless that this is not a
problem for a reasonable starting guess for the CMB power spectrum.

Standard numerical maximisation (minimisation) 
algorithms use differing amounts of gradient and/or curvature
information in order to arrive at the maximum point. 
For example, 
methods such as the downhill simplex or Powell's direction-set 
algorithm use no gradient information, requiring only function
evaluations (see Press et al. 1994). Other methods, such as the
variable metric technique, use only gradient information, whereas the
standard Newton--Raphson algorithm requires both gradient and
curvature information. In particular, Newton--Raphson technique 
and Powell's direction-set algorithm were investigated as methods of
numerically maximising the likelihood function.

\subsubsection{The Newton--Raphson method}
\label{nrmethod}

The Newton--Raphson method (or variations thereon) is the standard
numerical minimisation technique used in maximum-likelihood estimation
of the CMB power spectrum (e.g. Borrill 1999).  
Starting from some initial guess
$\bmath{a}_0$ for the flat band-powers, one performs the iteration
\begin{equation}
\bmath{a}_{n+1} = \bmath{a}_n + \delta\bmath{a}_n,
\label{eqn:iteration}
\end{equation}
where the correction $\delta\bmath{a}_n$ is given by
\[
\delta\bmath{a}_n = -[\bmath{H}^{-1}\bmath{g}]_{\bmath{a}=\bmath{a}_n},
\]
in which $\bmath{g}=\nabla(\ln{\cal L})$ is the gradient vector and 
$\bmath{H}=\nabla\nabla(\ln{\cal L})$ is the curvature (or Hessian) matrix.
The iteration (\ref{eqn:iteration}) 
is continued until the algorithm has converged to the
required accuracy. The elements of the gradient vector and curvature
matrix of the log-likelihood function (\ref{likedef2}) are easily shown to be
given respectively by
\begin{eqnarray}
\pd{\ln{\cal L}}{a_k}
& = & 
\tfrac{1}{2}\left[\bmath{v}^{\rm t}\bmath{C}^{-1}
\pd{\bmath{S}}{a_k}\bmath{C}^{-1}\bmath{v}-\mbox{Tr}\left(\bmath{C}^{-1}
\pd{\bmath{S}}{a_k}\right)\right], \label{eqn:loggrad}\\
\spd{\ln{\cal L}}{a_k}{a_k'}
& = & -\bmath{v}^{\rm t}
\bmath{C}^{-1}\pd{\bmath{S}}{a_k}
\bmath{C}^{-1}\pd{\bmath{S}}{a_{k'}}
\bmath{C}^{-1}\bmath{v} \nonumber \\
& &  \qquad\qquad 
+ \tfrac{1}{2}\mbox{Tr}\left(\bmath{C}^{-1}\pd{\bmath{S}}{a_k}
\bmath{C}^{-1}\pd{\bmath{S}}{a_{k'}}\right),\label{eqn:logcurv}
\end{eqnarray}
where $\mbox{Tr}(\,)$ denotes the trace of a matrix. 

The implementation of the Newton--Raphson algorithm can be divided
neatly into a series of computational steps (see Borrill 1999). For
the analysis of interferometer data, each iteration of the algorithm
consists of the following steps.
\begin{enumerate}
\item{Calculation of the signal covariance derivative matrices
$\partial\bmath{S}/\partial a_k$ $(k=1,\ldots,N)$, where $N$ is
the total number of parameters. In the flat band-power
parameterisation of the power spectra, we see from
(\ref{eqn:ss*2}) and (\ref{eqn:ss2}) that
$\bmath{S}$ is linear in the parameters $\bmath{a}$. Thus, the
derivative matrices are simply appropriate linear combinations
of the integrals $K_{ij}^{(a)}$ in (\ref{eqn:integrals}), 
which are evaluated once at the start of the calculation.}
\item{Construction of the (sparse) covariance matrix 
\[
\bmath{C} = \bmath{N}+\sum_{k=1}^N a_k\pd{\bmath{S}}{a_k},
\]
calculation of its (incomplete) Cholesky decomposition and solution of
the linear system $\bmath{Cx}=\bmath{v}$.}
\item{Solution of the linear systems
\[
\bmath{CY}_k= \pd{\bmath{S}}{a_k}\qquad (k=1,\ldots,N),
\]
where $\bmath{Y}_k$ is a matrix of the same dimensions as $\bmath{C}$.
Each column of $\bmath{Y}_k$ is solved for using the (incomplete)
Cholesky decomposition obtained in step (ii).}
\item{Assemble the elements of the gradient vector
\[
g_k \equiv \pd{\ln{\cal L}}{a_k} = \tfrac{1}{2}\left[
\bmath{v}^{\rm t}\bmath{Y}_k\bmath{x}-\mbox{Tr}(\bmath{Y}_k)\right].
\]
}
\item{Assemble the elements of the curvature matrix
\[
H_{kk'} \equiv \spd{\ln{\cal L}}{a_k}{a_{k'}}
=-\bmath{v}^{\rm t}\bmath{Y}_k\bmath{Y}_{k'}\bmath{x}+
\tfrac{1}{2}\mbox{Tr}(\bmath{Y}_k\bmath{Y}_{k'}).
\]
}
\item{Calculate the correction vector 
$\delta \bmath{a} = -\bmath{H}^{-1}\bmath{g}$.
}
\end{enumerate}

Although it is not necessary for the Newton--Raphson algorithm, we note that in
step (ii) one performs all the necessary calculations to evaluate the
log-likelihood function $\ln {\cal L}$ itself. This can be useful for
checking that its value is, in fact, decreasing with each iteration.
Typically, the Newton--Raphson algorithm converges to the maximum-likelihood
solution in approximately 5 iterations.

The computationally most demanding calculations in each iteration are
steps (ii) and (iii) above, which may be performed using either the
dense or sparse matrix techniques discussed in the previous section.
As mentioned in section~\ref{evalmulti}, for a visibility data vector
$\bmath{v}$ of length 5000, the
calculation in step (ii) of the (incomplete) Cholesky decomposition
requires approximately 5 min of CPU time for the dense matrix
{\sc lapack} routines and about 30 sec for the sparse matrix algorithm.  The
subsequent solution of the linear system $\bmath{Cx}=\bmath{v}$ then
requires about 0.25 sec or 0.5 sec respectively.  Thus, it would clearly
be faster to use the sparse matrix technique in step (ii).

In step (iii), however, one must solve a large number of linear
systems, namely one for each column of $\bmath{Y}_k$ for
$k=1,\ldots,N$, using the (incomplete) Cholesky decomposition
calculated in step (ii).  Even assuming the sky emission is due only
to the CMB, the total number of linear systems is $N_dN_b$, where
$N_d$ is the length of the data vector $\bmath{v}$ and $N_b$ is the
number of spectral bins in which one wishes to calculate the CMB power
spectrum. For a compact VSA observation of a single field, $N_d
\approx 5000$ and $N_b \approx 10$, and so one must solve $\sim 50000$
linear systems.  Solving each system requires around 0.5 sec of CPU time
using the sparse conjugate-gradient algorithm, but only 0.25 secs using
the {\sc lapack} dense matrix routines. Taking step (ii) and step
(iii) together, it is therefore faster to use the {\sc lapack}
routines, which require a total of about 3.5 hrs CPU time to 
complete these steps. The remaining
steps (iv) -- (vi) require only around 0.5 hr of CPU time, so that a
full Newton--Raphson 
iteration takes around 4 hrs.  Since 5 iterations of the 
Newton--Raphson algorithm 
are typically required for convergence, the total CPU to
calculate the maximum-likelihood solution is around 20 hrs.

We note that Bond, Jaffe \& Knox (1998, 2000) propose a variant of the 
Newton--Raphson algorithm, 
in which steps (v) and (vi) are different. In particular,
the correction vector in step (vi) is replaced by
\[
\delta\bmath{a} = -\bmath{F}^{-1}\bmath{g},
\]
where the {\em Fisher matrix} $\bmath{F}$ is the expectation value of 
the curvature matrix. From (\ref{eqn:logcurv}), the Fisher matrix has the
elements
\[
F_{kk'} \equiv \langle H_{kk'} \rangle = 
\tfrac{1}{2}\mbox{Tr}\left(\bmath{C}^{-1}\pd{\bmath{S}}{a_k}
\bmath{C}^{-1}\pd{\bmath{S}}{a_{k'}}\right).
\]
Thus, step (v) above is replaced by 
\begin{equation}
F_{kk'} = \tfrac{1}{2}\mbox{Tr}(\bmath{Y}_k\bmath{Y}_{k'}).
\label{eqn:fisher}
\end{equation}
Comparing the expressions for $H_{kk'}$ and $F_{kk'}$ we see that some
saving is made by using the latter. From (\ref{eqn:fisher}), however, we note
that the calculation of the Fisher matrix still requires knowledge of
the matrices $\bmath{Y}_k$ $(k=1,\ldots,N)$ calculated in step (iii),
which is the computationally most demanding part of the calculation.
Hence, the use of the Fisher matrix, rather than the full curvature
matrix, leads to a reduction of only around 20 mins in the CPU time
required for each Newton--Raphson iteration.

\subsubsection{Powell's direction-set method}
\label{powellmethod}

Clearly, the Newton--Raphson algorithm described above is computationally
intensive. Moreover, most time is spent is computing quantities
necessary for calculating the gradient and/or curvature of the
log-likelihood function. As discussed in section~\ref{sparsecg}, however, the
log-likelihood function itself can be calculated using the sparse
preconditioned conjugate-gradient method in about one-tenth of the CPU
time required by the standard {\sc lapack} dense matrix routines.
These observations suggest that it might be more computationally
efficient to maximise the log-likelihood function using an algorithm
that requires only function calls, rather than gradient and curvature
information.

A particularly efficient and robust estimator of this type is Powell's
direction-set method, which is described in detail by Press et al.
(1994). At each iteration, the algorithm determines a set of
`conjugate' directions at the corresponding point $\bmath{a}$ in the
parameter space, and minimises the function along each direction in
turn. The method can be shown to be quadratically convergent, so that
an exact quadratic form would be minimised in $N(N+1)$
line-minimisation, where $N$ is the dimensionality of the parameter
space. Typically, each line-minimisation requires 2 or 3 function
calls. Thus, one would expect between $2N(N+1)$ and $3N(N+1)$
functions calls would be required to minimise an exact quadratic form.

Since the algorithm requires only function evaluations, one may
straightforwardly employ the sparse matrix technique discussed in
section~\ref{sparsecg} for evaluating the log-likelihood function.  Assuming
the sky emission to be due solely to the CMB, the dimensionality of
the parameter space is simply the number of spectral bins in which one
wishes to estimate the CMB power spectrum, i.e.  $N=N_b$. For a
compact VSA observation of a single field, $N_b \approx 10$ and it was
found that the algorithm converged to the maximum-likelihood solution
after $\approx 300$ evaluations of the likelihood function. This is in
broad agreement with the anticipated number of function evaluations
quoted above, although the presence of the term $\ln |\bmath{C}|$ in
the log-likelihood function (\ref{likedef2}) means that it is not a simple
quadratic form. Thus, since each function evaluation requires
30 sec of CPU time, the maximum-likelihood CMB power spectrum may be obtained
in about 2.5 hrs of CPU time, which is considerably faster than the 
Newton--Raphson algorithm.

We note, in passing, that the downhill-simplex minimisation algorithm
(see Press et al. 1994) was also investigated. This again uses only
function calls, but it was found to converge too slowly. Nevertheless,
a robust technique for obtaining a accurate minimum in a reasonable
amount of CPU time is first to employ Powell's direction-set algorithm
until it converges to some point $\hat{\bmath{a}}$ in parameter space,
and then to restart the minimisation process from this point using a
downhill simplex. The final simplex step is then very fast and
provides a robust estimate of the minimum point. This algorithm
is implemented as the program {\tt maxlike} in the 
Microwave Anisotropy Dataset Computational sOftWare ({\sc madcow})
package (see section~\ref{sec:madcow}).

\subsubsection{Estimation of errors on parameter values}
\label{esterrors}

A major disadvantage of numerically maximising the likelihood is that
one remains ignorant of the global shape of the distribution. Therefore,
one cannot calculate marginalised distributions for each parameter,
in order to obtain proper confidence limits. Nevertheless, one
would hope that the shape of the likelihood function ${\cal L}$ near
its peak $\hat{\bmath{a}}$ should be well-approximated by a Gaussian,
so that
\begin{equation}
\ln{\cal L}(\hat{\bmath{a}}+\delta\bmath{a})
=  \ln{\cal L}(\hat{\bmath{a}}) + \tfrac{1}{2}\delta\bmath{a}^{\rm t}
\bmath{H}(\hat{\bmath{a}})\delta\bmath{a},
\label{eqn:gausslog}
\end{equation}
where $\bmath{H}$ is the curvature (or Hessian) matrix given in
(\ref{eqn:logcurv}), 
and the gradient term in the Taylor expansion vanishes since
$\hat{\bmath{a}}$ is a maximum. Exponentiating (\ref{eqn:gausslog}), we
immediately recognise that covariance matrix of our errors is given
simply by the inverse of the curvature matrix at the peak,
\[
\langle \delta\bmath{a}\,\delta\bmath{a}^{\rm t} \rangle
= \bmath{H}^{-1}(\hat{\bmath{a}}).
\]

The curvature matrix $H(\hat{\bmath{a}})$ may be calculated directly
by performing steps (ii), (iii) and (v) in the Newton--Raphson 
algorithm at the
maximum point $\hat{\bmath{a}}$. From section~\ref{nrmethod}, we note that,
for a compact VSA observation of a single field with 10 spectral bins,
this requires about 4 hrs of CPU time. However, we find that 
a computationally more efficient
procedure is to calculate the curvature matrix numerically by
performing second differences along each parameter direction. An
appropriate step size is easily chosen, and the algorithm requires
$\tfrac{1}{2}N(N+3)+1$ function evaluations. Thus, for 10 spectral
bins, one must evaluate the log-likelihood function around 70 times.  Using
the sparse preconditioned conjugate-gradient algorithm, this requires
only about 30 mins of CPU time. Comparing the
resulting curvature matrix with that obtained directly 
from (\ref{eqn:logcurv}),
we find that each element is agrees to within 0.1 per cent. The
numerical calculation of the curvature matrix and its inversion
to produce the errors covariance matrix is implemented in the
routine {\tt errors} in the {\sc madcow} package.

\subsection{Evaluation of the likelihood function along parameter axes}
\label{evalaxes}


The Powell direction-set minimiser, together with the sparse
preconditioned conjugate-gradient algorithm for evaluating the
log-likelihood function, provides a method of obtaining the 
maximum-likelihood power spectrum and the 
errors covariance matrix in a
reasonable amount of CPU time.  Nevertheless, by taking full advantage
of the sparse nature of the visibilities covariance matrix $\bmath{C}$
and of the flat band-power parameterisation of the power spectrum, it
is possible to calculate the maximum-likelihood solution in
considerably less computing time.

\begin{figure}
\begin{center}
\includegraphics[width=7cm]{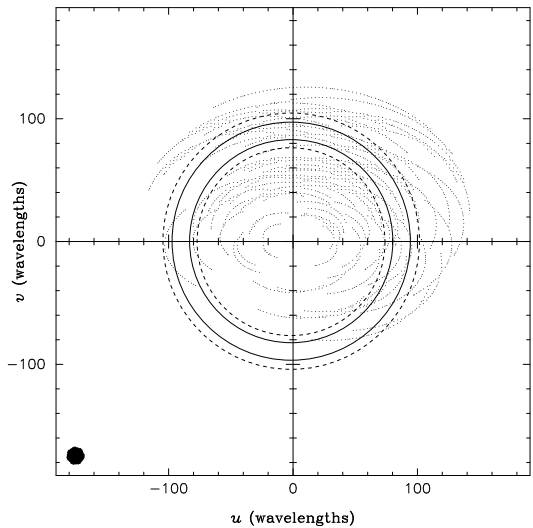}
\caption{An illustration of a typical annulus in the $uv$-plane, with
which the power spectrum is assumed constant (solid circles). This is
overlayed on the binned visibility samples shown in Fig.~\ref{fig2}.
The aperture function is plotted as a solid disc.
Only visibilities lying in the region between the two dashed circles are
sensitive to the level of the power spectrum in the annulus.}
\label{fig4}
\end{center}
\end{figure}

In the flat band-power parameterisation, one divides the observed
$\rho$-range into spectral bins of width $\Delta\rho$. This
corresponds to dividing the $uv$-plane into a set of concentric annuli
in each of which the power spectrum $D^{(p)}(\rho) = \rho^2
C^{(p)}(\rho,\nu_0)$ of each physical component in assumed constant,
and its amplitude is the parameter to be determined. A typical annulus
is illustrated by the solid circles in Fig.~\ref{fig4}, which are
overlayed on the visibilities samples shown in Fig.~\ref{fig2}. The extent
of the aperture function for a compact VSA observation of a single
field is plotted as the solid disc in the figure.

As discussed in section~\ref{visdata}, an interferometer visibility
measurement is only sensitive to the underlying sky Fourier modes
that lie within the aperture function at that point in the $uv$-plane.
Thus, only a fraction of the total number of measured visibilities
are sensitive to the amplitude of the power spectrum (of either the
CMB or some foreground component) in a particular annulus. This is
illustrated in Fig.~\ref{fig4}, where only the visibilities lying in the
region between the two dashed circles
are sensitive to the level of the power spectrum in
our typical annulus.

Thus, provided one is calculating the log-likelihood function along
the `axes' in parameter space, so that only one parameter value is changing
at any one time, the effective size of the dataset is much smaller
than the total number of data points. Thus, if the parameters 
$a_{k'\neq k}$ are fixed, 
\[
\ln{\cal L}(\bmath{v}|a_k) = \ln{\cal L}(\tilde{\bmath{v}}_k|a_k),
\]
where $\tilde{\bmath{v}}_k$ is the {\em reduced} data vector
consisting of only those visibilities sensitive to the 
flat band-power $a_k$. Given the geometry
illustrated in Fig.~\ref{fig4}, the length of the reduced data vector
will, in general, depend on the radius of the corresponding
$uv$-annulus, the extent of the aperture function and the positions of
the measured visibilities in the $uv$-plane.  For a compact VSA
observation of a single field, using $N_b=10$ spectral bins (or
$uv$-annuli), the reduced data vector typically contains around
3--10 times fewer data points than the full data vector.

The smaller size of the reduced data vector, as compared to the full
data vector, leads to a significant increase in the speed with which
the log-likelihood function can be calculated along the axes in
parameter space. For the standard {\sc lapack} dense matrix routines,
the cost of evaluating the log-likelihood function scales as 
$O(N_d^3)$,
where $N_d$ is the length of the data vector. Thus, by using the
reduced data vector, the computation time for evaluating the
log-likelihood function is reduced by a factor between 30 and 1000,
depending on the annulus in question.  

Unfortunately, the same
speed-up factor is not obtained for the sparse preconditioned
conjugate-gradient algorithm. This reason for this is clear from
Fig.~\ref{fig4}. Since the reduced data vector contains only visibilities
lying in the region between the two dashed circles, the corresponding
sparsity fraction $f_s$, will increase
significantly. Since the sparse matrix algorithm scales as
$O(f_s^{3/2}N_d^3)$, the log-likelihood evaluation is speed up only by a
factor between 4--60, depending on the annulus.  Nevertheless, for a
compact VSA observation of a single field, the sparse matrix approach
is still about a factor of 3 times faster in evaluating the
log-likelihood function than the {\sc lapack} 
dense matrix algorithm, when timings
are averaged over the 10 spectral bins.

\subsubsection{Calculation of the maximum-likelihood solution}
\label{mlsoln}

The speed-up outlined above is only possible when one varies only a
single parameter $a_k$, while keeping the others fixed.  Thus, one
cannot use a general minimisation algorithm such as the
Newton--Raphson or Powell direction-set method, since these perform
iterations of the form
\[
\bmath{a}_{n+1} = \bmath{a}_n + \delta\bmath{a}_n,
\]
where $\delta\bmath{a}_n$ may be in some general direction in
parameter space. A straightforward solution is provided, however,
simply by using a restricted version the Powell direction-set algorithm
in which the `conjugate' directions are forced to remain along the
parameter axes. Starting with some initial guess $\bmath{a}_0$ for the
solution, this procedure is equivalent to performing, in turn,
independent line-minimisations along each parameter axis, while
keeping the other parameters fixed at their initial values. One then
updates the whole solution vector and repeats the process until
convergence is
obtained. It was found that the likelihood function was sufficiently
structureless that this restricted `Powell' algorithm still converged
in around $3N(N+1)$ functions evaluations, where $N$ is the number of
parameters $\bmath{a}$.

For a compact VSA observation of a single field, with 10 spectral
bins, the maximum-likelihood power spectrum is obtained in about 10
mins of CPU time using the sparse matrix evaluation of the
log-likelihood function, or about 45 mins using standard {\sc lapack}
dense matrix routines. In both cases, this is
clearly faster than either of the methods discussed in section~\ref{nummax},
and is hence our method of choice for determining the
maximum-likelihood power spectrum from the current generation of CMB
interferometers.  The sparse and dense matrix versions of the
algorithm are implemented respectively in the programs
{\tt spfastlike} and {\tt lafastlike} in the {\sc madcow} package.

\subsubsection{Estimation of errors on parameter values}
\label{esterror2}

Once the maximum-likelihood solution $\hat{\bmath{a}}$ has been
obtained, it still remains to calculate the errors on our estimated
flat band-powers. One can, of course, simply calculate the curvature
matrix $\bmath{H}(\hat{\bmath{a}})$ numerically, in the manner
discussed in section~\ref{esterrors}, which requires about 30 mins 
of CPU time for
a compact VSA observation of a single field, with 10 spectral bins. 
This approach does, however, assume that the likelihood
function is well-approximated by a Gaussian near its peak, and leads to
symmetric error bars on our estimated parameter values. This may not,
in fact, be a valid approximation, especially for poorly constrained
parameters.

\begin{figure}
\begin{center}
\includegraphics[width=6.5cm]{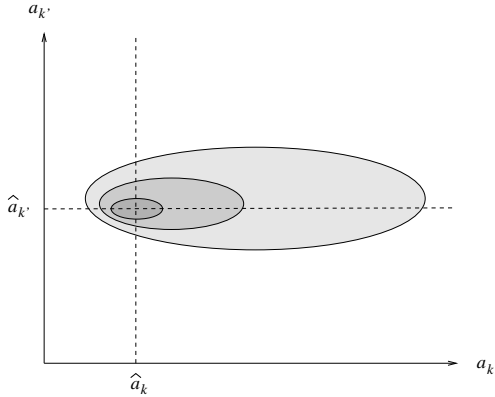}
\caption{An illustration of calculating the likelihood function
along each parameter direction, through the peak $\hat{\bmath{a}}$.}
\label{fig5}
\end{center}
\end{figure}

An alternative approach is to make use of the fact that it is usual to
choose the width $\Delta\rho$ of the spectral bins to correspond to
the characteristic width of the aperture function in the $uv$-plane.
Let us suppose that the sky emission is due only to the CMB.
In this case, all the flat band-power 
estimates $\hat{\bmath{a}}$ will be close to
uncorrelated. This is equivalent to the eigenvectors of
$\bmath{H}(\hat{\bmath{a}})$ lying close to the directions of the axes
in parameter space, and the off-diagonal elements being very small as
compared with those on the leading diagonal. In this case, a better
representation of the errors on our estimates is obtained by
evaluating the log-likelihood function along each parameter direction
in turn, through the point $\hat{\bmath{a}}$. Thus is illustrated in
Fig.~\ref{fig5}. In the ideal case, where the parameters $a_k$
$(k=1,\ldots,N)$ are independent, the resulting curves would be
the marginal distributions of each parameter.

In order to obtain accurate confidence intervals on each parameter, it
is best to calculate the log-likelihood function at a large number of
points along each parameter direction. Although, using the reduced
data vector approach discussed above, it is relatively cheap to
evaluate the log-likelihood function along each parameter direction
(while keeping the other parameters fixed), if one desires more than
around 100 points along each direction, a more efficient procedure is
to rotate to the signal-to-noise eigenbasis for each parameter $a_k$
in turn, as discussed in Appendix~\ref{sec:appndx:eigen}. Once this
rotation has been performed for a given parameter $a_k$, the
log-likelihood function can be calculated at a large number of points
along this parameter direction at virtually no additional computational
cost.

Rotating to the signal-to-noise eigenbasis requires the solution of a
generalised eigenproblem of dimension equal to the length of the
reduced data vector $\tilde{\bmath{v}}_k$ for each parameter. In the
absence of a stable sparse matrix implementation of this algorithm, a
standard {\sc lapack} subroutine was used to perform the calculation.
For a compact VSA observation of a single field, with 10 spectral
bins, the entire calculation requires about 30 mins of CPU time. This
approach is implemented as the routine {\tt eigenlike} in the {\sc
madcow} package.

In fact, the routine {\tt eigenlike} may be used in isolation to
obtain both the maximum-likelihood solution $\hat{\bmath{a}}$, and the
likelihood functions along each parameter direction passing through
the maximum. Starting with some initial guess $\bmath{a}_0$ one
performs a single `iteration' by calculating the likelihood functions
along the parameter directions through $\bmath{a}_0$. The peak
position of the likelihood in each direction is then used to update
the solution vector and the process is repeated until it converges.
For observations of a single field convergence is 
usually achieved after only 3 iterations. Thus, for a
compact VSA observation of a single field, with 10 spectral bins, the
full calculation requires about 1.5 hrs of CPU time. This is clearly
slower than using {\tt spfastlike} or {\tt lafastlike}, but provides
a useful cross check of one's results, since it uses entirely
different computational subroutines.

\subsubsection{Marginalisation over Galactic parameters}

For illustration, we assumed above  that the sky emission was due solely to CMB
fluctuations, without any contaminating Galactic foreground emission.
Let us now suppose instead that there is, in fact,
substantial foreground emission coming from (say) a single Galactic
component. If multifrequency observations are available,
one may calculate the maximum-likelihood solution $\hat{\bmath{a}}$ 
for the CMB and Galactic flat band-powers in each spectral bin,
using any of the techniques outlined above.

Suppose the
parameters $a_k$ and $a_{k'}$ correspond respectively to the CMB
and Galactic flat band-powers in the $k$th spectral bin. If the
bin width corresponds to the extent of the aperture function, these
two parameters will be quasi-uncorrelated with the remaining ones, but
the parameters $a_k$ and $a_{k'}$ themselves will be strongly
correlated since they refer to the same spectral bin.
Nevertheless, the routine {\tt eigenlike} provides a 
natural method for marginalising over
the flat band-power of the Galactic (or CMB) component in each bin.
Keeping the parameters $a_{k''}$ ($k'' \neq k$ and $k'' \neq k'$)
fixed, one can calculate the log-likelihood function in the
two-dimensional $(a_k,a_{k'})$-space by 
performing a signal-to-noise eigenmode rotation
at each required value of $a_{k'}$ (say), and calculating the variation
with respect to  $a_{k}$ at almost no extra computational cost.
One can then perform a numerical integration over
$a_k$ or $a_{k'}$, thereby obtaining a marginalised likelihood
function for the other parameter.

\subsection{Implementation: the {\sc madcow} package}
\label{sec:madcow}

The Microwave Anisotropy Dataset Computational sOftWare ({\sc madcow})
package consists of the 5 programs {\tt maxlike}, {\tt errors}, 
{\tt spfastlike}, {\tt lafastlike} and {\tt eigenlike}, which
are discussed above.  The first three programs are based on the sparse
matrix conjugate-gradient algorithm for calculating the log-likelihood
function, discussed in section~\ref{sparsecg}, whereas the last two programs
calculate the log-likelihood function using {\sc lapack} subroutines.

All the above programs have been parallelised using MPI, in order that
they may be run on distributed-memory parallel machines. The
parallelisation of the routines {\tt lafastlike} and {\tt
eigenlike} makes use of the {\sc S}ca{\sc lapack}
parallel version of the {\sc lapack} library. The speed of both 
routines is found to scale linearly with the number of processors up to
the maximum of 8 CPUs available on the Linux Beowulf cluster at our
disposal. The parallelisation of the spare matrix programs 
{\tt maxlike.c}, {\tt errors.c} and {\tt spfastlike.c} is less
straightforward, since existing parallel versions of the relevant
routines are not available. Some parallelisation of the algortihms
has been performed, but the resulting programs do not scale
as effectively as those using {\sc lapack} routines. A method of
achieving linear scaling of the sparse matrix algorithm is currently
under investigation. A $\beta$-test version of the {\sc madcow} package
may be downloaded from {\tt http://www.mrao.cam.ac.uk/software/madcow/}.

\section{Application to simulated observations}
\label{application}

\begin{figure}
\begin{center}
\includegraphics[angle=-90,width=7cm]{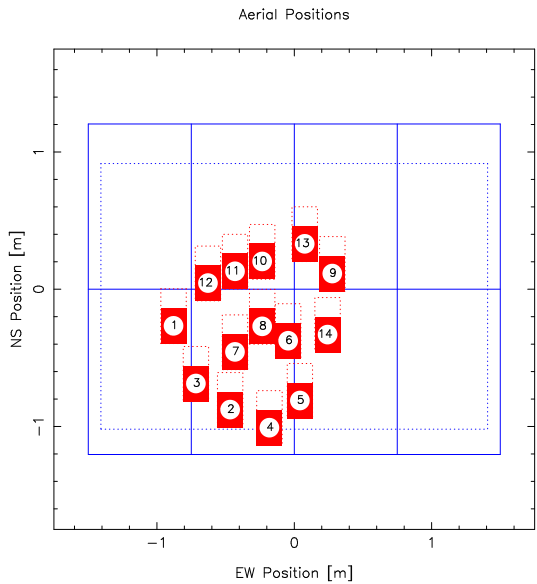}
\caption{The VSA compact array in June~2001. 
The rectangular area indicates the shape of the lower part of
the VSA table, which is mounted in east--west direction and
steerable in elevation.  There are 14~horn--reflector antennas
mounted at an inclination of 35~degrees. The dashed lines above
the horns show the extension of the antenna mount (Rusholme 2001). }
\label{fig:table}
\end{center}
\end{figure}

\begin{figure}
\begin{center}
\includegraphics[angle=-90,width=7.5cm]{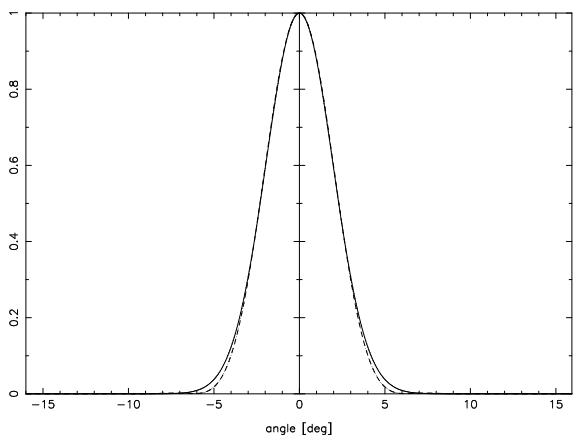}
\caption{The VSA primary (power) beam profile (dashed line) at 34.1~GHz
calculated by a ray-tracing program for the corrugated
horn-reflector antennas, compared to a Gaussian (solid line) of
the same FWHM (4.6 degrees).}
\label{fig:beam}
\end{center}
\end{figure}

As an illustration of our maximum-likelihood methods, in this section
we apply them to simulated VSA observations.  The VSA is a 14-element
radio interferometer, which is capable of operating at frequencies
between 26 and 36~GHz. The VSA is currently observing at a frequency
of 34.1~GHz. Its observing bandwidth is 1.5~GHz, and the system
temperature about 25-35~K. The telescope can observe in two different
configurations: the compact array uses corrugated horn-reflector
antennas of 14.3~cm diameter and baselines up to 1.5~m, and the
extended array consists of 32.2-cm horns with baselines up to about
3~m.  We base our simulations on the compact array with the actual
antenna arrangement as of June~2001, as shown in Fig.~\ref{fig:table}.
The primary (power) beam has a FWHM of 4.6 degrees at 34.1~GHz and can
be well approximated by a Gaussian (Fig.~\ref{fig:beam}).

We simulate VSA observations of three fields centred on ${\rm RA}
(1950)=0^{\rm h}~20^{\rm m}~00.0^{\rm s}$, ${\rm
  Dec.}~(1950)=30^{\circ}~00'~00''$, ${\rm RA} (1950)=0^{\rm
  h}~07^{\rm m}~20.6^{\rm s}$, ${\rm Dec.}~(1950)=30^{\circ}~16'~30''$
and ${\rm RA} (1950)=0^{\rm h}~12^{\rm m}~50.0^{\rm s}$, ${\rm
  Dec.}~(1950)=27^{\circ}~48'~00''$ (the VSA1, VSA1A and VSA1B fields
respectively). Real VSA observations of these fields (and others) have
been completed and are currently being analysed. The $uv$-coverage for
a 5~h-observation of the VSA1 field was shown in Fig.~\ref{fig1}(c).
The antenna arrangement yields a reasonably uniform sampling in the
multipole region $\ell \approx 100 - 900$.

\begin{figure}
\begin{center}
\includegraphics[angle=-90,width=8cm]{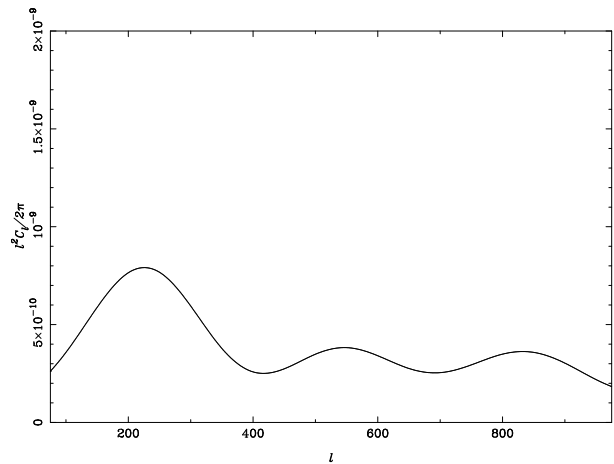}
\caption{The theoretical CDM power spectrum with parameter values
described in the text.}
\label{fig:trueps}
\end{center}
\end{figure}

In order to simulate an observation of CMB fluctuations, we first
create a $512\times 512$ array of Fourier modes, the real and
imaginary parts of which are drawn from a Gaussian distribution
with a variance given by
a theoretical power spectrum corresponding to 
a standard inflationary CDM model with
a Hubble parameter $H_0 = $~65~km~s$^{-1}$ Mpc$^{-1}$ and a fractional
baryon density $\Omega_{\rm b} = 0.05$.  The model is spatially-flat
with $\Omega_{\rm m}=0.3$ and $\Omega_\Lambda = 0.7$, a primordial
scalar spectral index $n = 1$ and no tensor modes. The spectrum is
normalised such that $Q_{\rm rms} = 18 \mu$K. The theoretical
$C_l$ spectrum is plotted in Fig.~\ref{fig:trueps}, and the 
actual realisation of the CMB
Fourier modes were plotted in Fig.~\ref{fig1}(a).

These Fourier modes are inverse Fourier-transformed to obtain the
corresponding CMB anisotropies on the sky. This array is then
multiplied by the telescope primary beam and Fourier transformed to
give a regular array in the visibility plane (see Fig.~\ref{fig1}(b)).
A VSA observation is then simulated by sampling the regular array at
the required points in the $uv$-plane using a bilinear interpolation.
The $uv$-positions at which the visibilities are sampled correspond to
a 64-s integration per visibility (see Fig.~\ref{fig1}(c)).

Instrumental noise on the data is simulated by adding to each
visibility a random complex number for which the real and imaginary
parts are drawn from a Gaussian distribution with an {\em rms} of
3.5~Jy. Since the measured VSA noise level of about 
240~Jy/$\sqrt{{\rm s}}$/baseline, this corresponds to $\sim$ 70 days of 5-h
observations. This is the time typically spent on each VSA field
during the first year of VSA observations between September~2000 and
August~2001.

These simulated VSA data typically contain $\approx 25\,000$ 64-s
visibility samples per day of observation. These visibilities
are binned as discussed in section~\ref{sec:binning}, with a
bin size of $\Delta u=3$~wavelengths. It was checked
that gridding with cell sizes $\Delta u
= 1-6$ did not significantly affect the final derived CMB power spectrum.
The binned data are then analysed using the maximum-likelihood techniques
described in section~\ref{sec:likeanalysis}. The results 
presented below were obtained using the 
program {\tt eigenlike} in the {\sc madcow} package.

We note that, for illustration, in these simulations 
we have restricted ourselves to the 
case in which the sky emission is due solely to
CMB fluctuations, without any 
contaminating foreground emission. 
Neglecting foregrounds provides a convenient way
of illustrating the general method, since only one observing frequency
is required, and one need not include the flat band-powers of the
Galactic power spectrum as parameters in our analysis.
Also, in practice, at the observing frequency of the VSA 
Galactic contribution can be largely ignored, because
the emission coming from Galactic dust and free-free radiation is
minimal. Moreover, VSA fields are chosen to lie in regions of low
Galactic emission. Nevertheless, as presented in 
section~\ref{sec:likeanalysis}, the maximum-likelihood methods
implemented in the {\sc madcow} package
can easily accommodate the presence of diffuse Galactic foregrounds.
Indeed, when multifrequency data are available, one may obtain
estimates of the power spectra of both the CMB and the Galactic
foreground emission.

\subsection{Single field observations}
\label{singlefield}

\begin{figure}
  \begin{center}
\includegraphics[angle=-90,width=8cm]{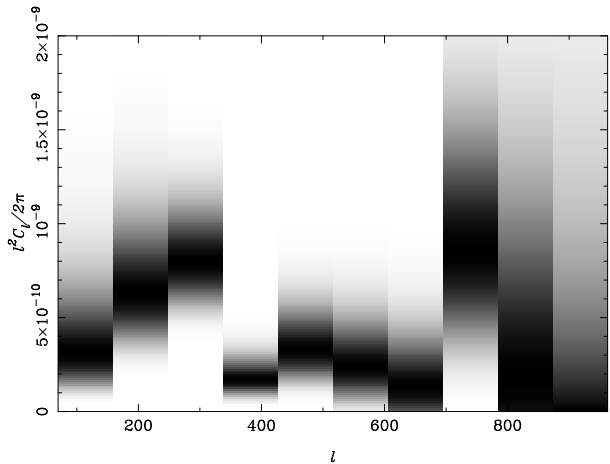}
    \caption{
      The maximum-likelihood CMB power spectrum derived from
simulated observation of the VSA1 field.
The flat band-powers were estimated in 10~spectral bins. The greyscale
bars show the likelihood function evaluated along the parameter
directions through the maximum-likelihood solution $\hat{\bmath{a}}$.}
    \label{fig:likeVSA1}
  \end{center}
\end{figure}

We first consider the case where the power spectrum is estimated from
only the simulated observations of the single field VSA1.
The flat 
band-powers have been estimated in 10 spectral bins of 
width $\Delta \ell \approx
90$ between $\ell \approx 80$ and $\ell \approx 950$. 
The resulting likelihood function 
evaluated along the parameter directions through
the maximum point $\hat{\bmath{a}}$ are plotted as the
greyscale bands in Fig.~\ref{fig:likeVSA1}. 

Dark areas indicate high likelihoods. The extension of the shaded
areas reflects the uncertainty in the estimated parameters. As
expected, the uncertainty is larger in the poorly sampled lower and
higher $\ell$--bins, where the telescope sensitivity is low. On the
other hand, the fairly well sampled multipoles around $\ell \approx
500$ can be determined more accurately. All reconstructed band powers
are consistent with the theoretical input power spectrum.
The bin width
was chosen to be equal to the $1/e$-diameter of the Fourier transform
of the primary beam. Thus, adjacent bins are only mildly correlated.
Indeed, on calculating the curvature matrix
$\bmath{H}(\hat{\bmath{a}})$ one finds that adjacent bins are
anti-correlated at only the 15 per cent level.

On repeating the
likelihood analysis for different realisations of the
simulated noise, it is found that results are always consistent
with the input spectrum and with one another.
It was further verified that approximating
the primary beam by a Gaussian of the same FWHM
does not distort the resulting power spectrum. 
Moreover, the
results were unaffected by ignoring
the finite bandwidth of the receivers and assuming a single observing
frequency in the centre of the band. 

\subsection{Mosaicing}
\label{sec:applic:mosaicing}

\begin{figure}
  \begin{center}
\includegraphics[angle=-90,width=8cm]{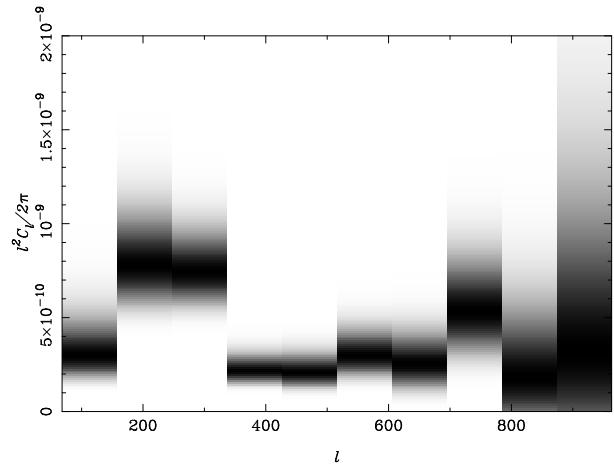}
    \caption{
      The maximum-likelihood CMB power spectrum derived from
simulated observation of the three mosaiced fields VSA1, VSA1A and
VSA1B.}
    \label{fig:likeVSA1AB}
  \end{center}
\end{figure}

We now turn to the simultaneous analysis of observations
of the three fields VSA1, VSA1A and VSA1B. Since these fields overlap
on the sky, the corresponding visibilities are thus correlated. 
The resulting likelihood function 
evaluated along the parameter directions through
the maximum point $\hat{\bmath{a}}$ are plotted as the
greyscale bands in Fig.~\ref{fig:likeVSA1AB}. 

Although mosaicing is
often used to improve the instrumental resolution in $\ell$-space, we
have chosen to compute the power spectrum in the same bins as for the
single field reconstruction, in order to facilitate a direct comparison. 
Moreover, owing to the relatively small size of the compact VSA horns, the
the aperture function is already sufficiently narrow to
provide the $\ell$-resolution necessary to resolve the expected
acoustic peak structure in the CMB power spectrum.  
As expected, the uncertainty on the flat band-powers 
has been significantly reduced,
particularly in the poorly sampled multipole regions. Moreover, the
anti-correlation between adjacent bins has reduced to around the 10 per
cent level.

\section{Conclusions}
\label{conc}

In this paper, we have investigated maximum-likelihood methods for
estimating the power spectrum of the cosmic microwave background (CMB)
from interferometer observations. In particular, we have considered
the computational efficiency of several techniques for obtaining
flat-band power estimates in a number of spectral bins,
together with confidence limits on the power in each bin. 
For multifrequency data, one may also estimate 
the flat band-powers of Galactic foreground emission in each spectral
bin. The methods developed may be applied to single-field or mosaiced
observations, and take proper account of non-coplanar baselines.

The sparse nature of the covariance matrix for
visibility data allows the use of sparse matrix techniques
which significantly reduce the computational burden of
evaluating the likelihood function, as compared
with standard dense matrix routines. This enables the maximum-likelihood
solution to be obtained more efficiently by numerically maximising the
likelihood using only function values, as opposed to more traditional
techniques, such as the Newton--Raphson algorithm, which rely 
on gradient and curvature
information. We also find the covariance matrix of the errors on
the parameters is most efficiently calculated using a numerical
second-differencing approach, rather than direct calculation of the
analytic expression for the curvature matrix. 

The speed with which the likelihood function is calculated may be
further increased by making use of the fact that only a small fraction
of the total number of observed visibilities are sensitive to the 
flat band-power in any one spectral bin. Using this reduced data-set
for each parameter enables very fast evaluation of the likelihood
function along the `axes' in parameter space. Indeed, we find that
performing independent line-maximisations of the likelihood function
along these parameter directions, and iterating until convergence,
provides the computationally most efficient way of obtaining the
maximum-likelihood solution.

If the spectral bins are chosen to be sufficiently wide that the flat
band-powers are quasi-uncorrelated parameters, one may dispense with
the Gaussian approximation to the likelihood function near its peak,
and obtain (generally asymmetric) confidence intervals on each
parameter by calculating the
likelihood function along each  parameter direction through the
maximum likelihood point. This is best achieved by performing
a single-to-noise eigenmode rotation for each parameter. Indeed, this
method can itself be used to arrive at the maximum-likelihood solution.
This is illustrated by application to simulated observations by 
the Very Small Array in its `compact' configuration.
If multifrequency data are available, this technique
may also be used to marginalise over the flat band-power of foreground
Galactic emission in each spectral bin.

\subsection*{ACKNOWLEDGMENTS}
KM acknowledges support from an EU Marie Curie Fellowship.

\onecolumn

\appendix

\section{The general form of the visibilities covariance matrix}
\label{sec:appndx:covariance}

The contribution ${\cal S}(\bmath{u},\nu)$ of the sky signal to the
measured visibility ${\cal V}(\bmath{u},\nu)$ is given simply by the
convolution of the underlying Fourier modes $a(\bmath{u},\nu)$ with
the aperture function $\tilde{A}(\bmath{u},\nu)$, which is itself the
Fourier transform of the primary beam $A(\bmath{x},\nu)$ of a single
antenna.  Assuming that the $j$th visibility is measured at the
position $\bmath{u}_j$ in the $uv$-plane and at a frequency $\nu_j$,
we may therefore write
\begin{equation}
{\cal S}_j \equiv S(\bmath{u}_j,\nu_j) = \int\!\!\!\int\!{\rm d}^2\bmath{u}
\,\tilde{A}(\bmath{u}_j-\bmath{u},\nu_j)a(\bmath{u},\nu_j).
\label{eqn:skyvis}
\end{equation}
If we write the functions $\tilde{A}(\bmath{u},\nu)$ and
$a(\bmath{u},\nu)$ in terms of their real and imaginary parts,
\[
\tilde{A}(\bmath{u},\nu)=\tilde{A}_R(\bmath{u},\nu)
+i\tilde{A}_I(\bmath{u},\nu),\qquad\qquad
a(\bmath{u},\nu)=a_R(\bmath{u},\nu)
+ia_I(\bmath{u},\nu),
\]
we quickly find that the real and imaginary parts of ${\cal S}_j$ are
given by
\begin{eqnarray}
{\cal S}^{(R)}_j & = & 
\int\!\!\!\int\!{\rm d}^2\bmath{u}\,
\left[
\tilde{A}_R(\bmath{u}_j-\bmath{u},\nu_j)a_R(\bmath{u},\nu_j)
-\tilde{A}_I(\bmath{u}_j-\bmath{u},\nu_j)a_I(\bmath{u},\nu_j)
\right], \nonumber\\
{\cal S}^{(I)}_j & = & 
\int\!\!\!\int\!{\rm d}^2\bmath{u}\,
\left[
\tilde{A}_R(\bmath{u}_j-\bmath{u},\nu_j)a_I(\bmath{u},\nu_j)
+\tilde{A}_I(\bmath{u}_j-\bmath{u},\nu_j)a_R(\bmath{u},\nu_j)
\right].
\label{eqn:sigvisri}
\end{eqnarray}
We note that if the antenna function $\tilde{A}(\bmath{x},\nu)$ is
real, then ${\cal S}_j^{(R)}$ depends only on the real parts
$a_R(\bmath{u},\nu)$ of the underlying Fourier modes of the sky, and
similarly ${\cal S}_j^{(I)}$ depends only on $a_I(\bmath{u},\nu)$.  If
the primary beam $A(\bmath{x},\nu)$ is real and even, then
$\tilde{A}(\bmath{x},\nu)$ is also real and even.

The elements of the constituent submatrices of the signal covariance
matrix $\bmath{S}$, introduced in
equation~(\ref{eqn:smatdef}), are most conveniently written
as
\begin{eqnarray}
S_{ij}^{(R,R)} 
\equiv \langle {\cal S}^{(R)}_i{\cal S}^{(R)}_j \rangle 
& = &
\tfrac{1}{2} [ \Re(\langle {\cal S}_i {\cal S}_j^\ast \rangle) +  
\Re (\langle {\cal S}_i {\cal S}_j\rangle) ],\nonumber\\
S_{ij}^{(I,I)} 
\equiv  \phantom{S}\langle {\cal S}^{(I)}_i{\cal S}^{I)}_j \rangle 
& = & 
\tfrac{1}{2} [ \Re(\langle {\cal S}_i {\cal S}_j^\ast \rangle) -  
\Re (\langle {\cal S}_i {\cal S}_j\rangle) ],\nonumber\\
S_{ij}^{(R,I)} 
\equiv  \phantom{S}\langle {\cal S}^{(R)}_i{\cal S}^{I)}_j \rangle 
& = & 
\tfrac{1}{2} [ -\Im(\langle {\cal S}_i {\cal S}_j^\ast \rangle) +  
\Im (\langle {\cal S}_i {\cal S}_j\rangle) ],\nonumber\\
S_{ij}^{(I,R)} 
\equiv  \phantom{S}\langle {\cal S}^{(I)}_i{\cal S}^{R)}_j \rangle 
& = & 
\tfrac{1}{2} [ \Im(\langle {\cal S}_i {\cal S}_j^\ast \rangle) +  
\Im (\langle {\cal S}_i {\cal S}_j\rangle) ],
\label{eqn:sigcovelem}
\end{eqnarray}
where $\Re$ and $\Im$ denote the real and imaginary parts of the, in
general, complex quantities $\langle {\cal S}_i{\cal S}_j^\ast\rangle$
and $\langle {\cal S}_i{\cal S}_j \rangle$, which are given by
\begin{eqnarray*}
\langle {\cal S}_i{\cal S}_j^\ast\rangle 
& = & 
\int\!\!\!\int\!{\rm d}^2\bmath{u}\,
\tilde{A}(\bmath{u}_i-\bmath{u},\nu_i)
\int\!\!\!\int\!{\rm d}^2\bmath{u}'\,
\tilde{A}^\ast(\bmath{u}_j-\bmath{u}',\nu_j)
\langle a(\bmath{u},\nu_i)a^\ast(\bmath{u}',\nu_j)\rangle \\
\langle {\cal S}_i{\cal S}_j\rangle, 
& = & 
\int\!\!\!\int\!{\rm d}^2\bmath{u}\,
\tilde{A}(\bmath{u}_i-\bmath{u},\nu_i)
\int\!\!\!\int\!{\rm d}^2\bmath{u}'\,
\tilde{A}(\bmath{u}_j-\bmath{u}',\nu_j)
\langle a(\bmath{u},\nu_i)a(\bmath{u}',\nu_j)\rangle.
\end{eqnarray*}
By definition, the matrix $\langle {\cal S}_i {\cal S}_j^\ast \rangle$
is Hermitian and $\langle {\cal S}_i {\cal S}_j \rangle$ is symmetric.
Similarly, each of the constituent submatrices $S^{(R,R)}_{ij}$ and
$S^{(I,I)}_{ij}$ is, by definition symmetric.  It is also
straightforward to show that $S^{(R,I)}_{ij} = S^{(I,R)}_{ji}$, so
that, as expected, the full covariance matrix $\bmath{S}$ is
symmetric. Moreover, from~(\ref{eqn:sigvisri}), we see that, if the aperture
function $\tilde{A}(\bmath{u},\nu)$ is real, and the real and
imaginary parts of the underlying Fourier modes $a(\bmath{u},\nu)$ are
uncorrelated, the elements of the off-diagonal submatrices
$S^{(R,I)}_{ij}$ and $S^{(I,R)}_{ij}$ are all zero. Thus, in this
case, the full covariance $\bmath{S}$ becomes block diagonal.

If the sky emission is due to $N_p$ physical components, then
\[
a(\bmath{u},\nu) = \sum_{p=1}^{N_p} a^{(p)}(\bmath{u},\nu).
\]
Moreover, if we assume the spectral index of each component does not
vary across the observed field, then we can describe the frequency
variation of each component through the functions $f_p(\nu)$
$(p=1,2,\ldots,N_p)$. In this paper, we take the reference frequency
$\nu_0=30$ GHz and normalise the frequency dependencies so that
$f_p(\nu_0)=1$ for all physical components.  If the emission from each
physical component is statistically isotropic over the observed field
and no correlations exist between the components, and using that fact
that the sky is real, the Fourier modes have mean zero and their
covariance properties are given by
\begin{eqnarray}
\langle a^{(p)}(\bmath{u},\nu) [a^{(p')}(\bmath{u}',\nu')]^\ast \rangle  
& = & C^{(p)}(\rho,\nu_0)f_p(\nu)f_{p'}(\nu') \delta_{pp'}
\delta(\bmath{u}-\bmath{u}'),\nonumber \\
\langle a^{(p)}(\bmath{u},\nu) a^{(p')}(\bmath{u}',\nu') \rangle  
& = & C^{(p)}(\rho,\nu_0)f_p(\nu)f_{p'}(\nu') \delta_{pp'}
\delta(\bmath{u}+\bmath{u}'),
\label{eqn:fouriercov}
\end{eqnarray}
where $\rho=|\bmath{u}|$ and $C^{(p)}(\rho,\nu_0)$ is the
ensemble-average power spectrum of the $p$th physical component at the
reference frequency $\nu_0$. If correlations are known {\em a priori}
to exist between physical components, then the expressions
(\ref{eqn:fouriercov}) can be straightforwardly generalised to include
them.  Assuming no correlations exist, however, we find that the
covariance structure of the Fourier modes of the total sky emission is
given by
\begin{eqnarray}
\langle {\cal S}_i {\cal S}_j^\ast \rangle&=&
\int\!\!\!\int\! \tilde{A} (\bmath{u}_i - \bmath{u}, \nu_i)
\tilde{A}^\ast (\bmath{u}_j - \bmath{u}, \nu_j) 
\xi(|\bmath{u}|,\nu_i,\nu_j)~{\rm d}^2\bmath{u}, 
\label{eqn:sigcorr1}\\
\langle {\cal S}_i {\cal S}_j\rangle&=&
\int\!\!\!\int\! \tilde{A} (\bmath{u}_i - \bmath{u}, \nu_i)
\tilde{A} (\bmath{u}_j + \bmath{u}, \nu_j) 
\xi(|\bmath{u}|,\nu_i,\nu_j)~{\rm d}^2\bmath{u},
\label{eqn:sigcorr2}
\end{eqnarray}
where we have introduced the {\em generalised power spectrum}
$\xi(\rho,\nu_i,\nu_j)$, which is defined by
\begin{equation}
\xi(\rho,\nu_i,\nu_j)=
\sum_{p=1}^{N_p} f_p(\nu_i)f_p(\nu_j)C^{(p)}(\rho,\nu_0).
\label{eqn:defgenps2}
\end{equation}
We note, in particular, that if the aperture function satisfies
$\tilde{A}^\ast(\bmath{u},\nu)=\tilde{A}(-\bmath{u},\nu)$, which
corresponds to a real primary beam $A(\bmath{x},\nu)$, then
\begin{equation}
\langle {\cal S}(\bmath{u}_i, \nu_i) {\cal S}_j (\bmath{u}_j,
\nu_j)\rangle=\langle {\cal S}(\bmath{u}_i, \nu_i) {\cal S}_j^\ast 
(-\bmath{u}_j,\nu_j)\rangle.
\label{eqn:equivcov}
\end{equation}
Thus, in this case, one need only evaluate $\langle {\cal S}_i{\cal
  S}_j\rangle$ and use the result~(\ref{eqn:equivcov}) in order to
construct the full signal covariance matrix $\bmath{S}$, whose
elements are defined in~(\ref{eqn:sigcovelem}).

If we introduce polar coordinates $(\rho,\theta)$ into the $uv$-plane,
we can write~(\ref{eqn:sigcorr1}) and~(\ref{eqn:sigcorr2}) formally as
the one-dimensional integrals
\begin{eqnarray}
\langle {\cal S}_i {\cal S}_j^\ast \rangle &=&
\int_0^\infty {\rm d}\rho\,\rho \,W^{(1)}_{ij}(\rho)
\,\xi(\rho,\nu_i,\nu_j),\nonumber \\
\langle {\cal S}_i {\cal S}_j\rangle&=&
\int_0^\infty {\rm d}\rho\,\rho \,W^{(2)}_{ij}(\rho)
\,\xi(\rho,\nu_i,\nu_j),
\label{eqn:corrwin}
\end{eqnarray}
where the {\em window functions} $W^{(1)}_{ij}(\rho)$ and
$W^{(2)}_{ij}(\rho)$ are defined respectively by
\begin{eqnarray}
W^{(1)}_{ij}(\rho)
& = & \int_0^{2\pi} {\rm d}\theta\, 
\tilde{A}(\bmath{u}_i-\bmath{u},\nu_i)
\tilde{A}^\ast (\bmath{u}_j-\bmath{u},\nu_j), \label{eqn:winfcn1}\\
W^{(2)}_{ij}(\rho)
& = & \int_0^{2\pi} {\rm d}\theta\, 
\tilde{A}(\bmath{u}_i-\bmath{u},\nu_i)
\tilde{A}(\bmath{u}_j+\bmath{u},\nu_j),\label{eqn:winfcn2}
\end{eqnarray}
and $\bmath{u} = \rho (\cos\theta,\sin\theta)^{\rm t}$.  In the case
where $\tilde{A}^*(\bmath{u},\nu)=\tilde{A}(-\bmath{u},\nu)$, one need
only evaluate $W^{(1)}_{ij}(\rho)$ and use (\ref{eqn:equivcov})
and~(\ref{eqn:sigcovelem}) to construct $\bmath{S}$.

\section{Mosaicing}
\label{sec:appndx:mosaicing}

For mosaiced observations, the pointing centres of individual
observations are not identical with the global phase--tracking centre.
This can be accommodated by shifting the primary beam response pattern
$A(\bmath{x}, \nu)$ to the corresponding pointing centre.  Thus, let
us assume that the $j$th visibility is measured at the position
$\bmath{u}_j$ in the $uv$-plane, at an observing frequency $\nu_j$,
and is referred to the pointing centre $\bmath{x}_j$.  In this case,
the expression corresponding to~(\ref{visdef}) becomes
\begin{eqnarray}
{\cal S}_j \equiv {\cal S}(\bmath{u}_j,\nu_j) &= &
\int\!\!\!\int\! A (\bmath{x} - \bmath{x}_j, \nu_j) \frac{\Delta T (\bmath{x}, \nu_j)}{T} e^{2 \pi i
  \bmath{u}_j \cdot \bmath{x}}~{\rm d}^2\bmath{x}\nonumber\\
&= &\int\!\!\!\int\! \tilde{A} (\bmath{u}_j - \bmath{u}, \nu_j)
e^{2\pi i (\bmath{u}_j-\bmath{u}) \cdot \bmath{x}_j} a(\bmath{u},
\nu_j)~{\rm d}^2\bmath{u}, 
\label{eqn:visconvshifted}
\end{eqnarray}
where the second equality follows from the Fourier shift theorem. This
illustrates that the effect of mosaicing is simply to introduce a
phase shift, so that the {\em effective} aperture function for the
$j$th field becomes
\begin{equation}
\tilde{A}_{\rm eff}(\bmath{u},\nu_j) 
=\tilde{A}(\bmath{u},\nu_j)\ 
e^{2\pi i \bmath{u} \cdot \bmath{x}_j},
\label{eqn:effapermos}
\end{equation}
which is, in general, complex.  The covariance properties of the
visibilities may then be obtained straightforwardly from the
expressions given in Appendix~\ref{sec:appndx:covariance}, by
replacing $\tilde{A}(\bmath{u},\nu)$ with $\tilde{A}_{\rm
  eff}(\bmath{u},\nu)$ given in~(\ref{eqn:effapermos}). In this case,
however, the corresponding window functions~(\ref{eqn:winfcn1})
and~(\ref{eqn:winfcn2}) will also be functions of $\bmath{x}_i$ and
$\bmath{x}_j$.  In particular, we find
\begin{eqnarray}
W^{(1)}_{ij}(\rho)
& = & 
e^{2\pi i (\bmath{u}_i \cdot \bmath{x}_i - \bmath{u}_j \cdot
\bmath{x}_j)} \int_0^{2\pi}{\rm d}\theta\, \tilde{A}
(\bmath{u}_i - \bmath{u},\nu_i) \tilde{A}^\ast(\bmath{u}_j -
\bmath{u},\nu_j) e^{2 \pi i \bmath{u}\cdot (\bmath{x}_j-\bmath{x}_i)},
\nonumber \\
W^{(2)}_{ij}(\rho)
& = & 
e^{2\pi i (\bmath{u}_i \cdot \bmath{x}_i + \bmath{u}_j \cdot
  \bmath{x}_j)} \int_0^{2\pi} {\rm d}\theta \,\tilde{A}
(\bmath{u}_i - \bmath{u},\nu_i) \tilde{A}(\bmath{u}_j +
\bmath{u},\nu_j) e^{2 \pi i \bmath{u}\cdot (\bmath{x}_j-\bmath{x}_i)}.
\end{eqnarray}
We note that, if the primary beam ${A}(\bmath{x},\nu)$ is real, then
the effective aperture function~(\ref{eqn:effapermos}) still satisfies
$\tilde{A}^\ast_{\rm eff}(\bmath{u},\nu) = \tilde{A}_{\rm
  eff}(-\bmath{u},\nu)$. Thus, (\ref{eqn:equivcov}) remains valid and one need
only evaluate $\langle {\cal S}_i{\cal S}_j^\ast\rangle$ (and hence
$W^{(1)}_{ij}(\rho)$) to construct $\bmath{S}$.

Although the above description of mosaicing is straightforward,
additional subtleties exist in practice, since interferometers such as
the VSA do not refer the phases of all visibilities to the same
constant phase-tracking centre.  Instead, the phases of visibilities
are referred to the pointing centre of the corresponding field. In this
case, the visibilities are given by
\begin{eqnarray}
{\cal S}_j \equiv {\cal S}(\bmath{u}_j,\nu_j) &= &
\int\!\!\!\int\! A (\bmath{x}, \nu_j) \frac{\Delta T (\bmath{x}+\bmath{x}_j, \nu_j)}{T} e^{2 \pi i
  \bmath{u}_j \cdot \bmath{x}}~{\rm d}^2 \bmath{x}\nonumber\\
&=&\int\!\!\!\int\! A (\bmath{x}-\bmath{x}_j, \nu_j)  \frac{\Delta T (\bmath{x}, \nu_j)}{T} e^{-2 \pi i \bmath{u}_j\cdot
  \bmath{x}_j} e^{2 \pi i
  \bmath{u}_j \cdot \bmath{x}}~{\rm d}^2 \bmath{x}\nonumber\\
&=& \int\!\!\!\int\! \tilde{A} (\bmath{u}_j-\bmath{u}, \nu_j)
 a(\bmath{u},
\nu_j) e^{-2\pi i \bmath{u} \cdot \bmath{x}_j}~{\rm
  d}^2\bmath{u}.
\label{eqn:visconvshifted2}
\end{eqnarray}
We note that (\ref{eqn:visconvshifted})
and~(\ref{eqn:visconvshifted2}) differ only by a phase shift
$\exp[-2\pi i \bmath{u}_j \cdot \bmath{x}_j]$. However, the vectors
$\bmath{u}_j$ are given in different coordinate systems, referring in
each case to the corresponding phase--tracking centre.
From~(\ref{eqn:visconvshifted2}), we see that we may obtain the
covariance properties of the corresponding visibilities from the
equations given in Appendix~\ref{sec:appndx:covariance}, provided we
replace $\tilde{A}(\bmath{u},\nu_j)$ by the new {\em effective}
aperture function for the $j$th field, which is given by
\begin{equation}
\tilde{A}_{\rm eff}(\bmath{u},\nu_j) = \tilde{A}(\bmath{u},\nu_j)\ 
e^{2\pi i(\bmath{u}-\bmath{u}_j) \cdot \bmath{x}_j},
\label{eqn:effapermos2}
\end{equation}
and similarly for the $i$ term. Hence, in this case, the window
functions are given by
\begin{eqnarray}
W^{(1)}_{ij}(\rho) 
& = &  \int_0^{2\pi} {\rm d} \theta\, \tilde{A}
(\bmath{u}_i - \bmath{u},\nu_i) \tilde{A}^\ast(\bmath{u}_j -
\bmath{u},\nu_j) e^{2 \pi i \bmath{u}\cdot (\bmath{x}_j-\bmath{x}_i)},
\nonumber \\
W^{(2)}_{ij}(\rho)
& = &  \int_0^{2\pi} {\rm d} \theta\, \tilde{A}
(\bmath{u}_i - \bmath{u},\nu_i) \tilde{A}(\bmath{u}_j +
\bmath{u},\nu_j) e^{2 \pi i \bmath{u}\cdot (\bmath{x}_j-\bmath{x}_i)},
\label{eqn:covvsa}
\end{eqnarray}
We note that, for a real primary beam, the relation
(\ref{eqn:equivcov}) again holds, so that only $\langle {\cal S}_i
{\cal S}_j^\ast \rangle$ (and hence $W^{(1)}_{ij}(\rho)$) need be calculated in
order to construct $\bmath{S}$.

\section{Large fields and non--coplanar baselines}
\label{sec:appndx:wcorrections}

In Appendices~\ref{sec:appndx:covariance} and
\ref{sec:appndx:mosaicing}, we have assumed that the field of view of
the telescope is small enough that the $w$-distortions discussed in
section~\ref{sec:wcorrections} can be neglected.  Without the
assumption of a small field size or vanishing $w$-components of the
baselines, the contribution of the sky signal to a measured visibility
is given by (e.g. Thompson et al.  1994\nocite{tms})
\begin{equation}
{\cal S} \equiv 
{\cal S}(\bmath{u}_j, w_j, \nu_j) =  \int\!\!\!\int\!{\rm d}^2\bmath{x}
\, A(\bmath{x},\nu_j) \frac{\Delta T}{T} (\bmath{x}, \nu_j) e^{2\pi i
  (\bmath{u}_j\cdot \bmath{x}+w_j (\sqrt{1-|\bmath{x}|^2}-1))}
\frac{1}{\sqrt{1-|\bmath{x}|^2}}, 
\end{equation}
where $w_j$ is the component of the baseline in the direction of the
phase--tracking centre. For convenience, we assume that pointing and
phase-tracking centres are identical. The factor
$\sqrt{1-|\bmath{x}|^2}-1 \approx -\frac{1}{2}|\bmath{x}|^2$ is small
(for example, $\frac{1}{2}|\bmath{x}|^2 < 0.001$ at an angle of
$x=2.3\degr$). We can thus either omit the factor
${1}/{\sqrt{1-|\bmath{x}|^2}} \approx 1$ or absorb it into the
definition of the primary beam $A(\bmath{x},\nu)$. We obtain
\begin{equation}
{\cal S}(\bmath{u}_j, w_j, \nu_j) =  \int\!\!\!\int\!{\rm d}^2\bmath{x}
\, A(\bmath{x},\nu_j) e^{-\pi i w_j |\bmath{x}|^2} \frac{\Delta T}{T}
(\bmath{x}, \nu_j) e^{2\pi i \bmath{u}_j\cdot \bmath{x}}.
\end{equation}
Defining an effective beam 
\begin{equation}
A_{\rm eff}(\bmath{x}, w, \nu_j) =
A(\bmath{x},\nu_j) \exp(-\pi i w |\bmath{x}|^2), 
\label{eqn:effbeamw}
\end{equation}
the visibilities are 
\begin{equation}
{\cal S}_j = {\cal S}(\bmath{u}_j, w_j, \nu_j) =  \int\!\!\!\int\!{\rm
  d}^2\bmath{u} 
\, \tilde{A}_{\rm eff}(\bmath{u}_j - \bmath{u},w_j, \nu_j) a(\bmath{u},
\nu_j), 
\end{equation}
where the Fourier transform of $A_{\rm eff}$ is given by a convolution
\begin{equation}
\tilde{A}_{\rm eff}(\bmath{u},w, \nu) = \int\!\!\!\int\!{\rm
  d}^2\bmath{u}'\, \tilde{A}(\bmath{u}-\bmath{u}', \nu) \frac{1}{i w}
  e^{i \pi |\bmath{u}'|^2/w}. 
\label{eqn:effaperw}
\end{equation}
The covariance structure of the visibilities can then by found by
using this expression in place of $\tilde{A}(\bmath{u},\nu)$ in the
results of Appendix~\ref{sec:appndx:covariance}. In particular, we note that if $\tilde{A}_{\rm
  eff} (\bmath{u}_i + \bmath{u}, w_i, \nu_i) = \tilde{A}_{\rm
  eff}^\ast ((-\bmath{u}_i) - \bmath{u}, -w_i, \nu_i)$, we find
\begin{equation}
 \langle {\cal S} (\bmath{u}_i, w_i,\nu_i) {\cal S} (\bmath{u}_j,
 w_j,\nu_j) \rangle = \langle {\cal S} (\bmath{u}_i, w_i,\nu_i) {\cal
 S}^\ast (-\bmath{u}_j, -w_j,\nu_j) \rangle. 
\label{eqn:equivcovw}
\end{equation}
Thus, in this case, the full signal covariance $\bmath{S}$ can be
obtained from~(\ref{eqn:sigcorr1}) and~(\ref{eqn:sigcovelem}).

\section{Gaussian primary beam}
\label{sec:appndx:gaussian}

Let us denote the primary beam at the $i$th observing frequency
$\nu_i$ by $A(\bmath{x},\nu_i)=\exp(-|\bmath{x}|^2/2\sigma_i^2)$,
where $\sigma_i$ is the dispersion of the primary beam at this
frequency.  The aperture function is simply the Fourier transform of
the primary beam, and is given by
$\tilde{A}(\bmath{u},\nu_i)=2\pi\sigma_i^2
\exp(-2\pi^2\sigma_i^2|\bmath{u}|^2)$. As we now show, in this case it
is possible to obtain analytical formulae for the window functions
$W^{(1)}_{ij}(\rho)$ and $W^{(2)}_{ij}(\rho)$.

\subsection{Two-dimensional approximation}
\label{sec:appndx:gaussian2d}

We begin by considering the general case of mosaiced observations, but
in the approximation that the visibilities can be obtained by a
two-dimensional Fourier transformation of the sky brightness
distribution and the primary beam. In that case,
from~(\ref{eqn:covvsa}), we have
\begin{equation}
W^{(1)}_{ij}(\rho)  = 
(2\pi\sigma_i\sigma_j)^2
\int\!\!\!\int\!{\rm d}^2\bmath{u}\,
\exp(-2\pi^2\sigma_i^2|\bmath{u}_i-\bmath{u}|^2)
\exp(-2\pi^2\sigma_j^2|\bmath{u}_j-\bmath{u}|^2)
\exp(2\pi i \bmath{u} \cdot (\bmath{x}_j - \bmath{x}_i)),
\label{eqn:winfcnmosinit}
\end{equation}
where all visibilities are referred to the pointing centre of the
corresponding field as in~(\ref{eqn:covvsa}). If we introduce the
vectors
\begin{equation}
\bmath{q}=\sigma_i^2\bmath{u}_i+\sigma_j^2\bmath{u}_j, \quad \bmath{x}
=\bmath{x}_j-\bmath{x}_i
\label{eqn:defwx}
\end{equation}
and use $\theta$ to denote the angle between $\bmath{q}$ and the
position vector $\bmath{u}$ in the Fourier plane, and $\phi$ the angle
between $\bmath{x}$ and $\bmath{q}$, we find that the window function
can be written as
\[
W^{1}_{ij} (\rho) = 
(2\pi\sigma_i\sigma_j)^2
\exp[-2\pi^2(\sigma^2_i|\bmath{u}_i|^2+\sigma^2_j|\bmath{u}_j|^2)]
\exp[-2\pi^2(\sigma_i^2+\sigma_j^2)\rho^2]
\int_0^{2\pi}\!{\rm d}\theta\,
\exp[4\pi^2|\bmath{q}|\rho\cos\theta]
\exp[2\pi i |\bmath{x}| \rho \cos(\theta+\phi)].
\]
The integral on $\theta$ can be written as
\begin{equation}
J_{ij} (\rho) 
=\int_0^{2\pi}\!{\rm d}\theta\,
\exp[4\pi^2|\bmath{q}|\rho\cos\theta]
\exp[2\pi i|\bmath{x}| \rho (\cos\theta\cos\phi-\sin\theta\sin\phi)]
=\int_0^{2\pi}\!{\rm d}\theta\,
\exp[p \cos\theta]
\exp[i a \cos\theta+i b \sin\theta],
\label{eqn:inttheta}
\end{equation}
where $p = 4\pi^2|\bmath{q}|\rho$, $a=2\pi |\bmath{x}| \rho \cos\phi$
and $b = -2\pi |\bmath{x}| \rho \sin\phi$ (compare Gradshteyn \&
Ryzhik\nocite{gradshteyn}, integrals 3.937). Substituting $z = \exp(i
\theta)$, ${\rm d}z = i z \,d\theta$, $\cos \theta = (z+z^{-1})/2$ and
$\sin \theta = (z-z^{-1})/(2i)$, we find
\[
J_{ij} (\rho) = \oint_{\cal C}\!{\rm d}z\,\frac{1}{iz}\exp\left[ (p + ia)
  \frac{z+z^{-1}}{2}+ ib\frac{z-z^{-1}}{2i}\right]
=  \oint_{\cal C}\!{\rm d}z\,\frac{1}{iz}\exp\left[ (p + b + ia)
  \frac{z}{2} + (p - b +ia) \frac{z^{-1}}{2}\right],
\]
where the curve ${\cal C}$ is the unit circle in the complex plane.
Defining $f(z) = z^{-1} \exp[ Az + Bz^{-1}]$, where $A =(p + b +
ia)/2$ and $B = (p-b+ia)/2$, we obtain
\[
J_{ij} (\rho) = \oint_{\cal C}\!{\rm d}z\,\frac{1}{i} f(z) = 2 \pi
\mbox{Res} (f,0),
\]
where $\mbox{Res} (f,0)$ denotes the residue of $f(z)$ around the
essential singularity $z=0$. The Laurent series of $f$ about this point
has the form $f(z)=\sum_{m=-\infty}^\infty c_m (z-0)^m$ and is given by
\begin{eqnarray*}
f(z) &=& \frac{1}{z} \exp[ Az + Bz^{-1}] = \frac{1}{z} \sum_{m=0}^\infty
\frac{(Az+Bz^{-1})^m}{m!}\\
&=& \frac{1}{z} \sum_{m=0}^\infty\sum_{n=0}^m
\frac{1}{m!} {^m}C_n A^n B^{m-n} z^{2n-m} = \sum_{m=0}^\infty\sum_{n=0}^m
\frac{1}{n!(m-n)!}  A^n B^{m-n} z^{2n-m-1}, 
\end{eqnarray*}
from which we can read off 
\[
\mbox{Res} (f,0) = c_{-1} = \sum_{n=0}^{\infty} \frac{(AB)^n}{(n!)^2}
= \sum_{n=0}^{\infty} \frac{(\frac{1}{2}\sqrt{4AB})^{2n}}{(n!)^2} = I_0 (\sqrt{4AB}), 
\]
where $I_0(z)=\sum_{n=0}^{\infty} (\frac{1}{2}z)^{2n} /(n!)^2$ is the
modified Bessel function of the first kind and order zero for a
complex argument~$z$.  Substituting all the abbreviations, we have
\begin{equation}
4AB  = p^2-a^2-b^2+2iap =
[(2\pi\rho)^2[(2\pi |\bmath{q}|)^2 - |\bmath{x}|^2   
  +i(4\pi \bmath{q} \cdot \bmath{x})].
\label{eqn:besselarg}
\end{equation}
Thus, the window function is given by the (complex) analytical
expression
\begin{equation}
W^{(1)}_{ij}(\rho) = 
(2\pi)^3(\sigma_i\sigma_j)^2
\exp[-2\pi^2(\sigma^2_i|\bmath{u}_i|^2+\sigma^2_j|\bmath{u}_j|^2)]
\exp[-2\pi^2(\sigma_i^2+\sigma_j^2)\rho^2]
I_0(2\pi \sqrt{k} \rho),
\label{eqn:winfcnmos}
\end{equation}
where
\begin{equation}
k=(2\pi|\bmath{q}|)^2-|\bmath{x}|^2+i(4\pi\bmath{q}\cdot\bmath{x}).
\label{eqn:besselarg2}
\end{equation}
The full signal covariance matrix $\bmath{S}$ can then be derived
using~(\ref{eqn:corrwin}) and~(\ref{eqn:sigcovelem}).

A useful check of our result~(\ref{eqn:winfcnmos}) is provided by the
special case of a single field, for which $\bmath{x}_i = \bmath{x}_j$,
and so $\bmath{x} = \bmath{x}_j - \bmath{x}_i = \bmath{0}$. Thus,
$J_{ij}(\rho)$ is given by the real integral
\[
J_{ij} (\rho) = 
\int_0^{2\pi}\!{\rm d}\theta\,
\exp[4\pi^2|\bmath{q}|\rho\cos\theta]
=\int_0^{2\pi}\!{\rm d}\theta\,
\exp[4\pi^2|\bmath{q}|\rho\sin\theta]
=2\pi I_0(4\pi^2|\bmath{q}|\rho).
\]
Therefore, in this case, the elements of the window function are given
by the real expression
\[
W^{(1)}_{ij}(\rho) =
(2\pi)^3(\sigma_i\sigma_j)^2
\exp[-2\pi^2(\sigma^2_i|\bmath{u}_i|^2+\sigma^2_j|\bmath{u}_j|^2)]
\exp[-2\pi^2(\sigma_i^2+\sigma_j^2)\rho^2]
I_0(4\pi^2|\sigma_i^2\bmath{u}_i+\sigma_j^2\bmath{u}_j|\rho),
\]
which is identical to setting $\bmath{x}=\bmath{0}$ in~(\ref{eqn:besselarg2}).
Comparison with~(\ref{eqn:sigcovelem}) shows that the offdiagonal blocks
$S_{ij}^{(I,R)}$ and $S_{ij}^{(R,I)}$ of the signal covariance matrix
vanish and real and imaginary parts of the visibilities can be treated
independently.

\subsection{Non-coplanar baselines and the $w$-correction}
\label{sec:appndx:gaussianw}

We now consider the effects of non-vanishing $w$-components on our
expression~(\ref{eqn:winfcnmos}) for the window function
$W^{(1)}_{ij}$.  From~(\ref{eqn:effbeamw}), for a Gaussian primary
beam of width $\sigma$, the effective primary beam is given by $A_{\rm
  eff}(\bmath{x}, w, \nu) = \exp(-|\bmath{x}|^2 (\frac{1}{2\sigma^2} +
i \pi w))$. On Fourier transforming, we obtain the effective aperture
function
\[
\tilde{A}_{\rm eff}(\bmath{u},w, \nu) = 2 \pi \sigma^2
\frac{1-i\phi}{1+\phi^2} e^{-2\pi^2\sigma^2 \frac{1-i\phi}{1+\phi^2}
  |\bmath{u}|^2} = 2 \pi \hat{\sigma}^2
(1-i\phi) e^{-2\pi^2\hat{\sigma}^2 (1-i\phi) |\bmath{u}|^2}, 
\]
where $\phi = 2 \pi \sigma^2 w$ and $\hat{\sigma}^2= \sigma^2 / (1 +
\phi^2) = \sigma^2 / (1 + 4 \pi^2 \sigma^4 w^2)$. 

We see that the effect of the $w$--distortion can be considered as
turning the primary beam into a complex Gaussian, whose Fourier
transform is again a complex Gaussian.  This can easily be taken into
account in the analysis from section~\ref{sec:appndx:gaussian2d} by
replacing
\[
\sigma^2 \rightarrow \sigma^2 \frac{1-i\phi}{1+\phi^2} =
\hat{\sigma}^2 (1 - i \phi) 
\]
in (\ref{eqn:winfcnmosinit}). For the angular integration
(\ref{eqn:inttheta}), we then have to use the vectors
\begin{equation}
\bmath{q}=\hat{\sigma}_i^2 \bmath{u}_i+\hat{\sigma}_j^2  \bmath{u}_j, 
\quad \bmath{x}
=\bmath{x}_j-\bmath{x}_i + 2 \pi \left(\phi_j \hat{\sigma}_j^2
\bmath{u}_j - \phi_i \hat{\sigma}_i^2 \bmath{u}_i
\right)
\label{eqn:newdefwx}
\end{equation}
instead of (\ref{eqn:defwx}).  The final expression for the window
function $W^{(1)}_{ij}(\rho)$ is
\begin{eqnarray*}
W^{(1)}_{ij}(\rho) & = &
(2\pi)^3(\hat{\sigma}_i \hat{\sigma}_j)^2 (1+\phi_i \phi_j + i (\phi_j
- \phi_i))\ 
e^{-2\pi^2(\hat{\sigma}^2_i|\bmath{u}_i|^2+\hat{\sigma}^2_j|\bmath{u}_j|^2)}
\ e^{-2\pi^2i (\phi_j \hat{\sigma}^2_j |\bmath{u}_j|^2 - \phi_i
\hat{\sigma}^2_i |\bmath{u}_i|^2)}\\
&&\times 
\ e^{-2\pi^2(\hat{\sigma}_i^2+\hat{\sigma}_j^2)\rho^2}
\ e^{-2\pi^2 i (\phi_j \hat{\sigma}_j^2- \phi_i \hat{\sigma}_i^2)\rho^2}
I_0(2\pi \sqrt{k} \rho),
\end{eqnarray*}
where now $k=(2\pi |\bmath{q}|)^2 - |\bmath{x}|^2 +i(4\pi \bmath{q}
\cdot \bmath{x})$ as in~(\ref{eqn:besselarg2}), but with $\bmath{q}$
and $\bmath{x}$ defined in (\ref{eqn:newdefwx}), and $\hat{\sigma}_i =
\sigma_i / (1+ \phi_i^2)$ with $\phi_i = 2 \pi \sigma^2_i w_i$.
Furthermore, we note that, $\tilde{A}_{\rm eff} (\bmath{u}_i +
\bmath{u}, w_i, \nu_i) = \tilde{A}_{\rm eff}^\ast ((-\bmath{u}_i) -
\bmath{u}, -w_i, \nu_i)$, and so~(\ref{eqn:equivcovw}) is satisfied.
Thus, the full signal covariance $\bmath{S}$ can be obtained
from~(\ref{eqn:equivcovw}) and~(\ref{eqn:sigcovelem}).

\section{Signal-to-noise eigenmodes}
\label{sec:appndx:eigen}

As discussed in section~\ref{esterror2}, it is useful to calculate the
log-likelihood function along each coordinate direction in parameter
space. In other words, one wishes to calculate the log-likelihood as a
function of a single parameter $a_k$, while the remaining parameters
are kept fixed; we denote this by $\ln{\cal L}(\bmath{v}|a_k)$. In fact, as
shown in section~\ref{evalaxes}, for interferometer data one may replace the
full data vector, $\bmath{v}$, with a reduced data vector
$\tilde{\bmath{v}}_k$ that contains only those visibilities sensitive
to changes in the flat band-power $a_k$. Hence, for various values of
$a_k$, one must evaluate
\[
\ln{\cal L}(\tilde{\bmath{v}}_k|a_k)
=\mbox{constant}-
\tfrac{1}{2}\left[\ln |\bmath{C}(\bmath{a})|
+\tilde{\bmath{v}}_k^{\rm t}\bmath{C}^{-1}(\bmath{a})
\tilde{\bmath{v}}_k\right].
\]
This calculation is neatly performed by using the method of signal-to-noise
eigenmodes, which are a special case of Karhunen-Loeve modes, where
the parameter dependence is linear (affine) (Bond 1995; Bond, Jaffe \&
Knox 1998; Tegmark, Taylor \& Heavens 1997).

The covariance matrix
$\bmath{C}(\bmath{a})=\bmath{S}(\bmath{a})+\bmath{N}$ contains
contributions from the sky signal and the noise. Since the parameters are
flat band-powers in spectral bins, we can write the signal
contribution as
\[
\bmath{S}(\bmath{a}) = \bmath{S}_1(a_k) +
\bmath{S}_2(\check{\bmath{a}}),
\]
where $\check{\bmath{a}}$ denotes collectively the fixed parameters
$a_{k'}$ $(k' \neq k)$. Hence, $\bmath{S}_1$ depends only on the
parameter of interest and $\bmath{S}_2$ depends only on the (fixed)
values of the other parameters, and is hence a fixed matrix. Moreover,
the matrix $\bmath{S}_1$ is linear in $a_k$, since it is simply a flat
band-power, and so we may write $\bmath{S}_1(a_k)=a_k\bmath{U}$, where
the fixed matrix $\bmath{U}=\bmath{S}_1(1)$. Thus the total covariance
matrix can be written as
\[
\bmath{C}(\bmath{a}) = a_k\bmath{U}+\bmath{S}_2(\check{\bmath{a}})
+ \bmath{N} \equiv a_k\bmath{U} + \bmath{V},
\]
where $\bmath{U}$ is the fixed `unit signal' covariance matrix and
$\bmath{V}$ is the fixed `generalised noise' matrix.

Since $\bmath{U}$ and $\bmath{V}$ are both real symmetric matrices,
they can be diagonalised simultaneously by a single similarity
transformation. This is most easily achieved by solving the generalised
eigenproblem
\[
\bmath{Ux}= \lambda\bmath{Vx}.
\]
Let us denote the corresponding eigenvalues by $\lambda_i$ and the
eigenvectors by $\bmath{e}_i$, which are normalised such that
$\bmath{e}_i^{\rm t}\bmath{V}\bmath{e}_i=1$. If we now define a new
set of `rotated' data by $\zeta_i = \bmath{e}_i\cdot\tilde{v}_k$, then
it is straightforward to show that these new variables are
uncorrelated for {\em any} value of the parameter $a_k$, and have a
covariance matrix with elements
\begin{equation}
\langle \zeta_i \zeta_j \rangle = (1+a_k\lambda_i)\delta_{ij}.
\label{eqn:uncorrl}
\end{equation}
The $\zeta_i$ called the signal-to-noise eigenmodes, and each is a
linear combination of the original data items. Modes corresponding to
a large eigenvalue are expected to be well-determined, while modes
with small eigenvalues are poorly-determined and do not contribute
significantly to the value of the likelihood function.

From (\ref{eqn:uncorrl}), we see that the log-likelihood, as a
function of the parameter $a_k$ (with the others held fixed) is given
simply by
\[
\ln{\cal L}(\bzeta|a_k)=\mbox{constant}-\tfrac{1}{2}
\sum_i
\left[\ln(1+a_k\lambda_i)+\frac{\zeta_i^2}{1+a_k\lambda_i}\right].
\]
This is trivial to evaluate and so the log-likelihood function can be
calculated at a large number of points along in $a_k$-direction at
virtually no extra computational cost.

\bsp 
\label{lastpage}

\begin{thebibliography}{}

\bibitem[\protect\citename{Anderson et al. }1999]{anderson}
Anderson E. et al., 1999, LAPACK Users' Guide, 3rd ed. SIAM, Philadelphia
\bibitem[\protect\citename{Baker et al. }1999]{baker}
Baker J.C. et al., 1999, MNRAS, 308, 1173 
\bibitem[\protect\citename{Bond }1995]{bond95}
Bond J.R., 1995, Phys.~Rev.~Lett., 74, 4369
\bibitem[\protect\citename{Bond et al. }1998]{bond98}
Bond J.R., Jaffe A.H., Knox L., 1998, Phys.~Rev.~D, 57, 2117
\bibitem[\protect\citename{Bond et al. }2000]{bond00}
Bond J.R., Jaffe A.H., Knox L., 2000, ApJ, 2000, 533, 19
\bibitem[\protect\citename{Borrill }1999]{borrill}
Borrill J., 1999, Proceedings of the 5th European SGI/Cray MPP
Workshop, astro-ph/9911389
\bibitem[\protect\citename{Cornwell }1988]{cornwell}
Cornwell T.J., 1988, A\&A, 202, 316
\bibitem[\protect\citename{Cornwell \& Perley }1992]{cornwellperley}
Cornwell T.J., Perley R.A., 1992, A\&A, 261, 353
\bibitem[\protect\citename{Cornwell et al. }1993]{cornwell93}
Cornwell T.J., Holdaway M.A., Uson J.M., 1993, A\&A, 271, 697 
\bibitem[\protect\citename{Christensen et al. }2001]{christensen}
Christensen N., Meyer R., Knox L., Luey B., 2001, Classical and
Quantum Gravity, 18, 2677
\bibitem[\protect\citename{Eaton }1983]{eaton}
Eaton M.L., 1983, Multivariate Statistics. Wiley, New York
\bibitem[\protect\citename{Ekers \& Rots }1979]{ekers}
Ekers R.D., Rots A.H., 1979, in van Schooneveld C., ed., ASSL
Vol. 76: IAU Colloq. 49: Image Formation from Coherence Functions in
Astronomy.  C.D. Reidel Publishing Co., p.~61
\bibitem[\protect\citename{Gradshteyn \& Ryzhik }1994]{gradshteyn}
Gradstheyn I.S., Ryzhik I.M., 1994, Table of Integrals, Series and
Products, 5th edition, Alan Jeffrey ed. Academic Press
\bibitem[\protect\citename{Greenbaum }1997]{greenbaum}
Greenbaum A., 1997, Iterative Methods for Solving Linear Systems. 
SIAM, Philadelphia
\bibitem[\protect\citename{Halverson et al. }2001]{halverson}
Halverson N.W. et al., 2001, astro-ph/0104489
\bibitem[\protect\citename{Hobson et al. }1995]{hobson}
Hobson M.P., Lasenby A.N., Jones M.E., 1995, MNRAS, 275, 863 
\bibitem[\protect\citename{Hobson \& Magueijo }1996]{hobsonmagueijo}
Hobson M.P., Magueijo J., 1996, MNRAS, 283, 1133
\bibitem[\protect\citename{Holdaway }1999]{holdaway}
Holdaway M.A., 1999, in ASP Conf. Ser. 180: Synthesis Imaging in
Radio Astronomy II, p.~401
\bibitem[\protect\citename{Jones }1990]{jones}
Jones M.E., 1990, Ph.D. Thesis, University of Cambridge
\bibitem[\protect\citename{Jones }1997]{jones97}
Jones M.E., 1997, in Bouchet F.R., Gispert R. et al, eds., Microwave
Background Anisotropies, XVIth Moriond Astrophysics Meeting. Editions
Frontieres, p.~161
\bibitem[\protect\citename{Jones \& Scott }1998]{jonesscott}
Jones M.E., Scott P.F., 1998, in Tr\^{a}n Thanh V\^{a}n J. et al.,
eds., Proceedings of XXXIIIrd Rencontres de Moriond, Fundamental
  Parameters in Cosmology, Editions Frontieres, p.~233
\bibitem[\protect\citename{Leitch et al. }2001]{leitch}
Leitch E.M. et al., 2001, astro-ph/0104488
\bibitem[\protect\citename{Maisinger et al. }1997]{maisinger}
Maisinger K., Hobson M. P., Lasenby A. N., 1997, MNRAS, 290, 313
\bibitem[\protect\citename{Mather et al. }1994]{mather}
Mather J.C. et al., 1994, ApJ, 420, 439
\bibitem[\protect\citename{O'Sullivan et al.  }1994]{osullivan}
O'Sullivan C. et al., 1995, MNRAS, 274, 861
\bibitem[\protect\citename{Padin et al. }2001]{padin}
Padin S. et al., 2001, ApJ, 549, L1 
\bibitem[\protect\citename{Pearson et al. }2000]{pearson}
Pearson T.J. et al., 2000, IAU Symposium 201, E2
\bibitem[\protect\citename{Perley }1999]{perley}
Perley R.A., 1999, in {\em Synthesis Imaging in Radio Astronomy II},
eds. Taylor G. B. et al., ASP Conference Series, Vol. 180, p.~383
\bibitem[\protect\citename{Press et al. }1994]{press}
Press W.H., Teukolsky S.A., Vetterling W.T., Flannery B.P., 1994, 
Numerical Recipies in Fortran 77, CUP, Cambridge
\bibitem[\protect\citename{Rusholme }2001]{rusholme}
Rusholme B., 2001, Ph.D. Thesis, University of Cambridge
\bibitem[\protect\citename{Scott et al. }1996]{scott}
Scott P.F. et al., 1996, ApJ, 461, L1
\bibitem[\protect\citename{Smoot et al. }1992]{smoot}
Smoot G. et al., 1992, ApJ, 396, L1
\bibitem[\protect\citename{Taylor }2000]{taylor00}
Taylor A.C., 2000, IAU Symposium 201, E5
\bibitem[\protect\citename{Taylor et al.}2001]{taylor01}
Taylor A.C., Grainge K., Jones M.E., Pooley G.G., Saunders R.,
Waldram E.M., 2001, MNRAS, 327, L1 
\bibitem[\protect\citename{Tegmark et al. }1997]{tegmark}
Tegmark M., Taylor A.N., Heavens A.F., 1997, ApJ, 480, 22
\bibitem[\protect\citename{Thompson, Moran \& Swenson }1994]{tms}
Thompson A. R., Moran J. M., Swenson G. W., 1994, Interferometry and
Synthesis in Radio Astronomy. Krieger Publishing Company, New York
\bibitem[\protect\citename{White et al. }1999a]{white}
White M., Carlstrom J.E., Dragovan M., Holzapfel W.L., 1999, ApJ,
514, 12 
\bibitem[\protect\citename{White et al. }1999b]{whiteb}
White M., Carlstrom J., Dragovan M., Holzapfel S.W.L., 1999,
astro-ph/9912422  


\end{thebibliography}
\end{document}